\documentclass[times,twocolumn,final]{elsarticle}

\usepackage{medima}

\usepackage{outline}
\usepackage{amssymb}
\usepackage{amsmath}
\usepackage{graphicx}
\usepackage{times}
\usepackage{bm} 
\usepackage{url}

\usepackage{epstopdf}

\usepackage{epsfig}
\usepackage{tikz}
\usetikzlibrary{spy}
\usepackage{algpseudocode}
\usepackage{algorithm}
\usepackage{mathrsfs}

\usepackage{nicefrac}       
\usepackage{booktabs}       

\usepackage{amsmath,graphicx,caption,mathtools,amssymb,amsthm}
\usepackage{graphicx}
\usepackage{multirow}
\usepackage{subcaption}
\usepackage{accents}

\newtheorem{proposition}{\bf{Proposition}}

\def\code#1{\texttt{#1}}

\long\def\comment#1{} 

\newcommand{\xmath}[1] {\ensuremath{#1}\xspace}
\newcommand{\blmath}[1] {\xmath{\bm{#1}}}

\newcommand{\x}{\blmath{x}}

\newcommand{\0}{\blmath{0}}

\newcommand{\norm}[1] {\xmath{\left\| #1 \right\|}}

\newcommand{\Ib}{{\blmath I}}

\newcommand{\fb}{{\blmath f}}

\renewcommand{\sb}{{\blmath s}}
\newcommand{\tb}{{\blmath t}}

\newcommand{\wb}{{\blmath w}}
\newcommand{\xb}{{\blmath x}}
\newcommand{\yb}{{\blmath y}}
\newcommand{\zb}{{\blmath z}}

\newcommand{\Nc}{\mathcal{N}}
\newcommand{\Sc}{\mathcal{S}}

\newcommand{\Rd}{{\mathbb R}}
\newcommand{\Cd}{{\mathbb C}}

\newcommand{\thetab}{{\boldsymbol {\theta}}}

\newcommand{\Pc}{{{\mathcal P}}}

\newcommand{\Ed}{{{\mathbb E}}}

\newcommand{\beq}{\begin{equation}}
\newcommand{\eeq}{\end{equation}}
\newcommand{\beqa}{\begin{eqnarray}}
\newcommand{\eeqa}{\end{eqnarray}}
\newcommand{\Fc}{{\mathcal F}}

\usepackage{amsmath}
\usepackage{framed,multirow}
\usepackage{amssymb}
\usepackage{latexsym}
\usepackage{url}
\usepackage{xcolor}
\usepackage{xspace}
\usepackage{adjustbox}
\usepackage{makecell}
\usepackage{tabularx}
\usepackage{algorithm}
\usepackage{tikz}
\usetikzlibrary{decorations.pathreplacing,calc}
\newcommand{\tikzmark}[1]{\tikz[overlay,remember picture] \node (#1) {};}
\newcommand*{\AddNote}[4]{%
    \begin{tikzpicture}[overlay, remember picture]
        \draw [decoration={brace,amplitude=0.5em},decorate,ultra thick,black]
            ($(#3)!(#1.north)!($(#3)-(0,1)$)$) --  
            ($(#3)!(#2.south)!($(#3)-(0,1)$)$)
                node [align=center, text width=2.0cm, pos=0.5, anchor=west] {#4};
    \end{tikzpicture}
}%
\newcommand*{\AddNoteb}[4]{%
    \begin{tikzpicture}[overlay, remember picture]
        \draw [decoration={brace,amplitude=0.5em},decorate,ultra thick,black]
            ($(#3)!(#1.north)!($(#3)-(0,1)$)$) --  
            ($(#3)!(#2.south)!($(#3)-(0,1)$)$)
                node [align=center, text width=2.0cm, pos=0.5, anchor=west] {#4};
    \end{tikzpicture}
}%

\newcolumntype{C}{ >{\centering\arraybackslash m{16mm}}}
\newcolumntype{M}[1]{>{\centering\arraybackslash}m{#1}}
\usepackage{hyperref}

\definecolor{newcolor}{rgb}{.8,.349,.1}

\journal{Medical Image Analysis}

\begin{document}

\verso{Chung and Ye}

\begin{frontmatter}

\title{Score-based diffusion models for accelerated MRI}

\author[1]{Hyungjin Chung}
\author[1]{Jong Chul Ye\corref{cor1}}
\cortext[cor1]{Corresponding author.}
\ead{jong.ye@kaist.ac.kr}

\address[1]{Department of Bio and Brain Engineering, Korea Advanced Institute of Science and Technology (KAIST), Daejeon 34141, Republic of Korea}

\received{?}
\finalform{?}
\accepted{?}
\availableonline{?}
\communicated{?}

\begin{abstract}
Score-based diffusion models provide a powerful way to model images using the gradient of the data distribution. Leveraging the learned score function as a prior, here we introduce a way to sample data from a conditional distribution given the measurements, such that the model can be readily used for solving inverse problems in imaging, especially for accelerated MRI. In short, we train a continuous time-dependent score function with denoising score matching. Then, at the inference stage, we iterate between the numerical SDE solver and data consistency  step to achieve reconstruction. Our model requires magnitude images only for training, and yet is able to reconstruct complex-valued data, and even extends to parallel imaging. The proposed method is agnostic to sub-sampling patterns and has excellent generalization capability so that it can be used with any sampling schemes for any body parts that are not used for training data. Also, due to its generative nature, our approach can quantify uncertainty, which is not possible with standard regression settings. On top of all the advantages, our method also has very strong performance, even beating the models trained with full supervision. With extensive experiments, we verify the superiority of our method in terms of quality and practicality. Code available at: \href{this repository}{https://github.com/HJ-harry/score-MRI} 
\end{abstract}


\end{frontmatter}


\section{Introduction}
\label{sec: intro}

\begin{figure*}[!hbt]
    \centering\includegraphics[width=18cm]{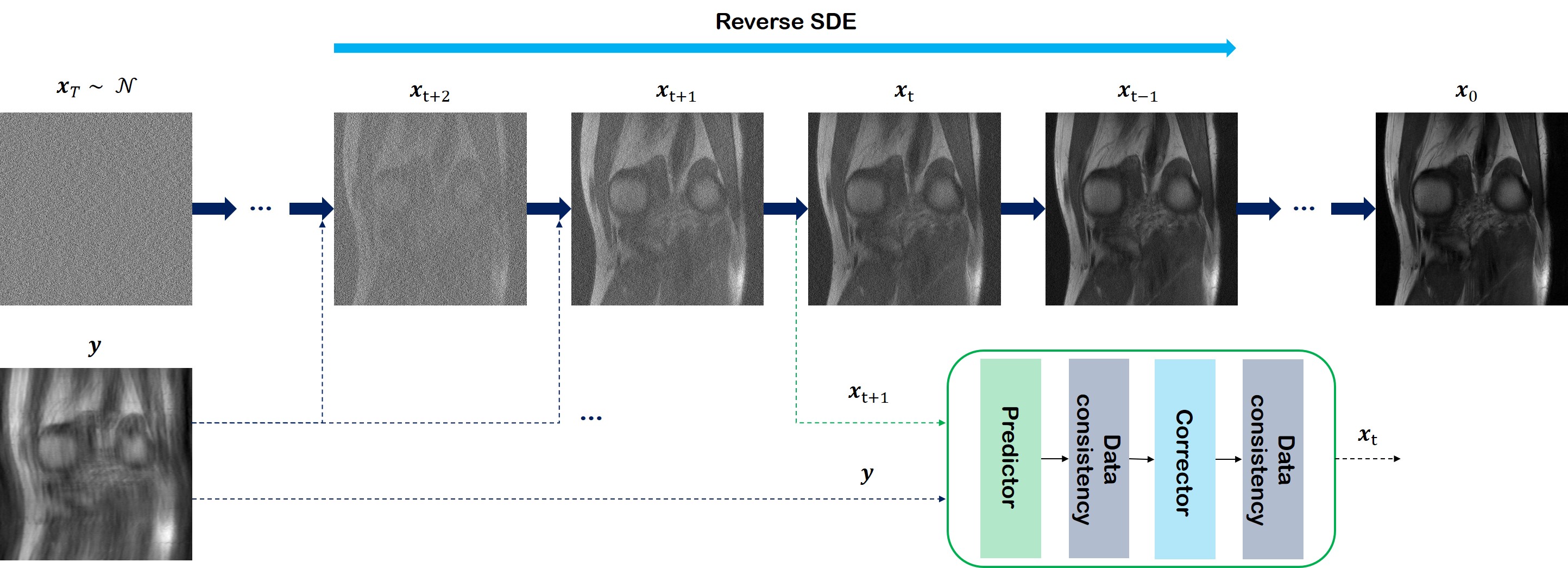}
    \caption{Overview of the proposed method. Starting from $\xb_T$, sampled from the prior distribution, $\xb_0$ is reached by solving the reverse SDE with score-based sampling, alternating between the update step, and the data consistency step.}
	\label{fig:concept}
\end{figure*}

Reconstruction methods from sub-sampled measurements for magnetic resonance imaging (MRI) have seen a lot of progress over the years. Regularized reconstruction methods exploit the sparsity of hand-crafted priors along with data consistency to arrive at a solution~\citep{donoho2006compressed}, yet the construction of priors is often non-trivial, and {\em none} of them can properly model the actual complicated data distribution $p_{\text{data}}$ of MRI scans. Data-driven deep learning methods train the models either directly~\citep{wang2016accelerating} or indirectly~\citep{oh2020unpaired, yaman2020self}, yet the methods rely heavily on the supervision of a well-curated large database of $k$-space data, which is hard to obtain.

Recently, score-based models~\citep{hyvarinen2005estimation, song2019generative}, and denoising diffusion probabilistic models (DDPMs)
~\citep{sohl2015deep, ho2020denoising} have gained wide interest as a new class of generative model that achieves surprisingly high sample quality without adversarial training~\citep{song2020score, nichol2021improved, dhariwal2021diffusion}. Among many works, \citet{song2020score} generalized discrete score-matching procedures to a continuous stochastic differential equation (SDE), which in fact also subsumes diffusion models into the same framework. We refer to score-based SDEs as score-based diffusion models henceforth to emphasize that our proposed methodology can be flexibly used with any realizations from the two model classes: score-based generative models, and diffusion models.

Score-based diffusion models perturb the data distribution according to the forward SDE by injecting Gaussian noise, arriving at a tractable distribution (e.g. isotropic Gaussian distribution). In order to sample from the data distribution, one can train a neural network to estimate the gradient of the log data distribution (i.e. score function, $\nabla_\xb \log p(\xb)$), and use it to solve the reverse SDE numerically. Unconditional generation of samples from $p(\xb)$ using these score-based diffusion models have found their applications in image~\citep{song2020score, nichol2021improved, dhariwal2021diffusion}, audio~\citep{kong2020diffwave}, and even graph~\citep{niu2020permutation} synthesis. Conditional generation from $p(\xb|\yb)$ has also been studied in the context of widely known computer vision problems: in-painting~\citep{song2019generative, song2020score}, super-resolution~\citep{choi2021ilvr, saharia2021image}, and image editing~\citep{meng2021sdedit}.

In this work, we propose a framework generally applicable to solving inverse problems in imaging and especially focus on the task of MRI reconstruction. 
Notably, our method requires training a {\em single} score function with {\em magnitude} images only.
With the trained  score model at hand using the de-noising score matching loss, we construct a solver for the reverse SDE from the variance exploding (VE)-SDE~\citep{song2020score}, which enables us to sample from the distribution $p(\xb|\yb)$, conditioned on the measurement $\yb$. This is done by imposing  data consistency step at every iteration, after the unconditional update step.

Despite the fact that the score function was not trained to solve the image reconstruction task, our method achieves state-of-the-art performance, even outperforming models that were trained in a supervised fashion specifically for image reconstruction tasks for complex images. Furthermore, our model is agnostic to the sub-sampling pattern used in the acceleration procedure, as opposed to supervised models which require re-training every time a new sampling scheme is designed. In addition, the proposed method can be extended to the reconstruction of complex-valued MR image acquisition using a {\em single} network that has {\em never} seen complex-valued data before.
Even more, our method can be readily applied to practical multi-coil settings with the same score function, where the update via score function can be applied in parallel to each coil image. 
 It is worth mentioning that the generalization capability of the trained score function is far greater. In fact, when we try the reconstruction of data that is heavily out of training data distribution (e.g. different contrast, and even different anatomy), we are still able to achieve high fidelity reconstruction.

Finally, the proposed method is inherently stochastic, and for that, we can sample multiple reconstruction results from the same measurement vector $\yb$. This is especially useful since we are able to quantify uncertainty without any specific treatment (e.g. Monte Carlo dropout~\citep{gal2016dropout}, estimating variance directly~\citep{kendall2017uncertainties}) to the neural net. We observe that at low acceleration factors, posterior samples do not deviate much from each other, meaning high confidence of the network. As the acceleration factor is pushed to higher values, the variance gradually increases, which can potentially aid practitioners' decision-making.

Given the amazing generalization capability and the flexibility to produce uncertainty maps, we believe that the  proposed score-based approach can be an important framework for compressed sensing MRI.
 Overview of the proposed method is illustrated in Fig.~\ref{fig:concept}, and we detail our proposed method in Section~\ref{sec:main_contributions}.

The manuscript is organized as follows: relevant background on score-based diffusion models is reviewed in Section~\ref{sec:background}; detailed procedure and algorithm of the proposed method is presented in Section~\ref{sec:main_contributions}; specifications about the implementation and experiments are given in Section~\ref{sec:methods}; experimental results are featured in Section~\ref{sec:results}; discussion about the concurrent works of conditional generation using diffusion models, limitations, and broader impacts of our work is presented in Section~\ref{sec:discussion}; we conclude our work in Section~\ref{sec:conclusion}.

\section{Background}
\label{sec:background}

\subsection{Score-based SDE}

One can construct a continuous diffusion process $\{\xb(t)\}_{t=0}^T$ with $\xb(t)\in \Rd^n$, where $t \in [0, T]$ is the time index of the progression
and $n$ denotes the image dimension. We choose $\xb(0) \sim p_{data}$ and $\xb(T) \sim p_T$, where $p_{data}, p_T$ refers to the data distribution of interest, and the prior distribution which are tractable to sample from (e.g. spherical Gaussian distribution), respectively. Then, the stochastic process can be constructed as the solution to the following SDE
\begin{equation}\label{eq:forward_SDE}
    d\xb = \fb(\xb, t)dt + g(t)d\wb,
\end{equation}
where $\fb : \Rd^n \mapsto \Rd^n$ and $g: \Rd \mapsto \Rd$ correspond to the drift coefficient, and the diffusion coefficient, respectively, and $\wb$ is a standard $n-$dimensional Brownian motion.

One can construct different SDEs by choosing different functions for $\fb$ and $g$. First, by choosing 
\begin{equation}\label{eq:VP-SDE}
    \fb = -\frac{1}{2}\beta(t)\xb,\quad g=\sqrt{\beta(t)},
\end{equation}
where $0 < \beta(t) < 1$ is a monotonically increasing function of noise scale, one achieves the variance preserving (VP)-SDE \citep{ho2020denoising}. In this case, the magnitude of the signal decays to 0, and the variance is preserved to a fixed constant as $t \rightarrow \infty$. In fact, VP-SDE can be seen as the continuous version of DDPM~\citep{song2020score, kingma2021variational}. Therefore, while DDPM was developed in a separate variational framework, it can also be seen as a realization of SDE.

On the other hand, variance exploding (VE) SDEs choose 
\begin{equation}\label{eq:VE-SDE}
\fb = \0,\quad g = \sqrt{\frac{d[\sigma^2(t)]}{dt}},
\end{equation}
where $\sigma(t) > 0$ is again a monotonically increasing function, typically chosen to be a geometric series~\citep{song2019generative, song2020score}. Unlike VP-SDE, VE-SDE diffuses the signal with a very large variance, which explodes as $t \rightarrow \infty$, hence its name. Empirically, we found that using VE-SDE typically leads to higher sample qualities, and hence focus on developing our method on top of VE-SDE hereafter. However, we note that the use of VP-SDE (including the family of DDPMs developed under the variational framework) is also straightforward under our framework.

Interestingly, the reverse process of \eqref{eq:forward_SDE} can be constructed with another stochastic process \citep{song2020score}:
\begin{align}\label{eq:reverse_SDE}
    d\xb &= [\fb(\xb, t) - g(t)^2 \underbrace{\nabla_\xb \log p_t(\xb)}_{\text{score function}}]dt + g(t) d\bar{\wb} \\
    &= \frac{d[\sigma^2(t)]}{dt} \underbrace{\nabla_\xb \log p_t(\xb)}_{\text{score function}} + \sqrt{\frac{d[\sigma^2(t)]}{dt}} d\bar{\wb} \notag,
\end{align}
where $dt$ is the infinitesimal {\em negative} time step, and $\bar{\wb}$ is again the standard $n-$dimensional Brownian motion running backwards. The last equality follows by plugging in Eq.~\eqref{eq:VE-SDE} to Eq.~\eqref{eq:reverse_SDE}. 

In order to solve \eqref{eq:reverse_SDE}, one has to know the score function for all $t$. One can estimate this score function with a time-conditional neural network $s_{\thetab}(\xb(t), t) \simeq \nabla_{\xb} \log p_t(\xb(t))$, and replace the term in \eqref{eq:reverse_SDE}. Since we do not know the \textit{true} score, we can instead use denoising score matching~\citep{vincent2011connection}, where we replace the unknown $\nabla_\xb \log p_t(\xb)$ with $\nabla_\xb \log p_{0t}(\xb(t)|\xb(0))$, where $p_{0t}(\xb(t) | \xb(0))$ is the Gaussian perturbation kernel which perturbs the probability density $p_0(\xb)$ to $p_t(\xb)$. Under some regularity conditions, $s_\theta$ trained with denoising score matching will satisfy $s_{\theta^{*}}(\xb(t), t) = \nabla_\xb \log p_t(\xb)$ almost surely~\citep{song2020sliced}.
Formally, we optimize the parameters $\thetab$ of the score network with the following cost:
\begin{align}\label{eq:score_cost}
    \min_{\thetab} \Ed_{t \sim U(0, 1)} \Big[ \lambda(t) &\Ed_{\xb(0)} \Ed_{\xb(t)|\xb(0)}\Big[ \\ \notag
    &\norm{\sb_{\thetab} (\xb(t), t) - \nabla_{\xb} \log p_{0t} (\xb(t) | \xb(0))}_2^2 \Big]
    \Big],
\end{align}
where $\lambda(t)$ is an appropriate weighting function, e.g. likelihood weighting of~\cite{song2021maximum}, which puts different emphasis according to the time $t$.
In the case of Gaussian perturbation kernels, the gradient of the perturbation kernel can be formulated explicitly: $\nabla_{\xb} \log p_{0t} (\xb(t) | \xb(0)) = (\xb(t) - \xb(0))/\sigma(t)^2$.
Intuitively, one can also understand \eqref{eq:score_cost} as training the neural network to de-noise $\xb(t)$, which was constructed by adding noise to $\xb(0)$. 

In \cite{song2019generative}, it was shown with an illustrative toy example that when you add Gaussian noise to the random variable, you essentially get a blurred version of the original density, which indeed comes from the property that the addition of two random variables corresponds to the convolution of two densities~\citep{loeve2017probability}. Hence, when the score function is trained to denoise the given data across multiple noise scales, one can start the diffusion process with pure noise and gradually decrease the noise following the gradient information of the data density. Consequently, one can arrive at high-density modes of the data distribution.

Once the network is trained with \eqref{eq:score_cost}, we can plug the approximation $s_{\thetab}(\xb, t) \simeq \nabla_{\xb} \log p_t(\xb(t))$ to solve the reverse SDE in Eq. \eqref{eq:reverse_SDE}:
\begin{equation}
    d\xb = \frac{d[\sigma^2(t)]}{dt} s_{\thetab}(\xb(t), t) + \sqrt{\frac{d[\sigma^2(t)]}{dt}} d\bar{\wb}.
\end{equation}
Then, we can solve the SDE numerically, for example, with Euler-Maruyama discretization~\citep{song2020score}. This involves discretizing $t$ in range [0, 1] uniformly into $N$ intervals such that $0 = t_0 < t_1 < \dots < t_N = 1$, with $\Delta t = 1 / N$. Additionally, we can correct the direction of gradient ascent with {\em corrector} algorithms such as Langevin MC~\citep{parisi1981correlation}. Iteratively applying predictor and corrector steps yield the predictor-corrector (PC) sampling algorithm~\citep{song2020score}, as presented in Algorithm~\ref{alg:PC}. With the algorithm presented in Algorithm~\ref{alg:PC}, we can sample from the distribution $p(\xb)$. In Section~\ref{sec:main_contributions}, we extend this sampling scheme to a conditional sampling algorithm, which enables us to sample from $p(\xb|\yb)$.

\begin{algorithm}[!hbt]
\caption{Predictor-Corrector (PC) sampling}
\begin{algorithmic}[1]
\Require $\sb_{\theta}, N, M, \{\epsilon_i\}$ \Comment{step size}, $\{\sigma_i\}$ \Comment{noise schedule}
\State $\xb_N \sim \Nc({\bf 0}, \sigma_{T}^{2} \Ib)$
\For{$i = N-1:0$} \do \\
\State $\xb'_i \gets \xb_{i+1} + (\sigma_{i+1}^2 - \sigma_{i}^2) \sb_{\theta}(\xb_{i+1}, \sigma_{i+1})$\tikzmark{top}  \tikzmark{right}
\State $\zb \sim \Nc(\bf{0}, \bf{I})$
\State $\xb_i \gets \xb'_i + \sqrt{\sigma_{i+1}^2 - \sigma_i^2}\zb$\tikzmark{bottom}
\For{$j = 1:M$} \do \\
\State $\zb \sim \Nc(\bf{0}, \bf{I})$\tikzmark{top2}
\State $\xb'_i \gets \xb_i + \epsilon_{i} \sb_\theta (\xb_i, \sigma_i)$
\State $\xb_i \gets \xb'_i + \sqrt{2\epsilon_{i}}\zb$\tikzmark{bottom2}
\EndFor
\EndFor
\State \textbf{return} $\xb_0$
\end{algorithmic}\label{alg:PC}
\AddNote{top}{bottom}{right}{Predictor}
\AddNoteb{top2}{bottom2}{right}{Corrector}
\end{algorithm}

\section{Main Contributions}
\label{sec:main_contributions}

\subsection{Forward Measurement Model}

In accelerated MRI, we consider the following measurement model
\begin{align}\label{eq:forward_model}
    \yb = A\xb 
\end{align}
where $\yb \in \Cd^{m}$ is the measurement, $\xb \in \Cd^{n}$ is the latent image,  and a parameterized forward measurement matrix $A \in \Cd^{m \times n}$ is defined as
\begin{equation}\label{eq:mm_mri}
    A := \Pc_{\Omega} \Fc \Sc,
\end{equation}
where $\Sc := [\Sc^{(1)}; \dots; \Sc^{(c)}]$ is the sensitivity map for $c$ different coils, $\Fc$ denotes Fourier transform, and $\Pc_\Omega$ is a diagonal
matrix with zeros and ones that represent the sub-sampling operator with the given sampling pattern $\Omega$. The sensitivity map $\Sc$ are normalized such that we have
\begin{equation}\label{eq:normalization}
     \Sc^* \Sc = I .
\end{equation}
In the case of single-coil acquisition, $\Sc$ reduces to identity matrix such that $A_{(sc)} = \Pc_{\Omega} \Fc$.

\subsection{Reverse SDE for Accelerated MR Reconstruction}

A classic approach to find the solution of Eq.~\eqref{eq:forward_model} is to solve the following constrained optimization problem:
\begin{eqnarray}\label{eq:pls}
    \min_{\xb}& \Psi(\xb) \\
    \mbox{subject to} & \yb = A\xb, \notag
\end{eqnarray}
where $\Psi(\cdot)$ is, for example, a sparsity promoting regularizer derived from compressed sensing (CS) theory~\citep{donoho2006compressed}, such as $\ell_1$ wavelet~\citep{lustig2007sparse} and total variation (TV)~\citep{block2007undersampled}.
Solving Eq.~\eqref{eq:pls} typically involves proximal algorithms such as variable splitting~\citep{boyd2011distributed} or
projection onto the convex sets (POCS) \citep{samsonov2004pocsense}, which decouples the optimization of the prior term, and the forward consistency term. Then, one can alternate between solving the two sub-problems to arrive at the optimum.

In Bayesian perspective, we immediately see that $\Psi(\xb)$ in Eq.~\eqref{eq:pls} is the prior model of the data, i.e. $p(\xb)$. 
 Hence, we can imagine that by more accurately estimating the complex prior data distribution, one would be able to achieve higher quality samples.

That being said, one of the important  differences of the proposed method compared to the classical approaches is that rather than modeling the prior distribution $p(\xb)$, we exploit its stochastic samples. Specifically,
the samples from the prior distribution can be obtained from the reverse SDE in Eq.~\eqref{eq:reverse_SDE},
which  can be discretized  as illustrated in Algorithm~\ref{alg:PC} with
\begin{align}\label{eq:pr1}
    \xb_{i} \gets (\sigma_{i+1}^2 - \sigma_{i}^2)\sb_{\theta}(\xb_{i+1}, \sigma_{i+1})
    + \sqrt{\sigma_{i+1}^2 - \sigma_{i}^2}\zb,
\end{align}
Then, the data consistency mapping on the constraint in \eqref{eq:pls}
can be implemented by
\begin{align}\label{eq:pr2}
     \xb_i \gets \xb_i + \lambda A^*(\yb - A\xb_i) = (I-\lambda A^*A)\xb_i + A^*\yb
\end{align}
for $\lambda \in [0,1]$, where $A^*$ denotes the Hermitian adjoint of $A.$. 
 
 Similar to our companion work \citep{chung2021come}, we
 impose the constraint on the operator $A$ such that $ (I-\lambda A^*A)$ is a {\em non-expansive mapping} \citep{bauschke2011convex}:
\begin{align}
\|(I-\lambda A^*A)\x-(I-\lambda A^*A)\x'\|\leq  \|\x-\x'\|,\quad \forall \x,\x'
\end{align}
For example, 
 projection onto convex sets (POCS) in \citep{tang2011projection, fan2017projections} 
or the
one-iteration of the standard gradient descent with controlled step size~\citep{jalal2021robust, ramzi2020denoising}
corresponds to the non-expansive data consistency mapping. 
In the following,  the normalization step in \eqref{eq:normalization} is shown essential to ensure
that $(I-\lambda A^*A)$ is indeed non-expansive:
\begin{proposition}
With the sensitivity normalization in \eqref{eq:normalization}, $(I-\lambda A^*A)$  is non-expansive for $\lambda \in [0,1]$.
\end{proposition}
\begin{proof}
Using the properties of the spectral norm, we have
\begin{align*}
\|A^*A\| &= \|\Sc^* \Fc^* \Pc_{\Omega}^*\Pc_{\Omega} \Fc \Sc\| \stackrel{\text{(a)}}{=}  \|\Sc^* \Fc^* \Pc_{\Omega} \Fc \Sc\| \\
 & \stackrel{\text{(b)}}{\leq}   \|\Sc^* \Fc^* \Fc \Sc\|  \stackrel{\text{(c)}}{=}  \|\Sc^* \Sc\|  \stackrel{\text{(d)}}{=} 1
\end{align*}
where (a)(b) come that the subsampling operator $P_\Omega$ is a diagonal matrix with 0 and 1,
(c) is from the orthonormality of the Fourier transform and (d) is from \eqref{eq:normalization}.
Therefore, we have
\begin{align*}
\|I-\lambda A^*A\|&\leq \max\{|1-\lambda|, 1\}\leq 1 
\end{align*}
for $\lambda \in [0,1]$.  Accordingly,
\begin{align}
\|(I-\lambda A^*A)\x-(I-\lambda A^*A)\x'\|\leq  \|I-\lambda A^*A\| \|\x-\x'\| \leq \|\x-\x'\| 
\end{align}
This concludes the proof.
\end{proof}

 Application of Eqs.~\eqref{eq:pr1} and \eqref{eq:pr2} correspond to the predictor step.
When using the additional corrector steps as in Algorithm~\ref{alg:PC}, one can also apply the same treatment to the discrete corrector step
\begin{align}\label{eq:corrector-POCS}
    \xb_{i} &\gets \xb_{i+1} + \epsilon_i \sb_{\theta}(\xb_{i+1}, \sigma_{i+1})
    + \sqrt{2\epsilon_{i}}\zb \\ \notag
    \xb_i &\gets \xb_i + \lambda A^{*}(y - A\xb_i),
\end{align}
where $\epsilon_i$ is the step size at the $i^{\text{th}}$ iteration.
Iteratively applying predictor and corrector steps as in PC algorithm gives rise to the inference algorithm, which is described formally in Algorithm~\ref{alg:PC-POCS} when
$\lambda =1$. 

\begin{algorithm}[!t]
\caption{Score-based sampling (Real)}
\begin{algorithmic}[1]
\Require $\sb_{\theta}, N, M, \{\epsilon_i\}$ \Comment{step size}, $\{\sigma_i\}$ \Comment{noise schedule}
\textbf{Define} $A := \Pc_{\Omega} \Fc$
\State $\xb_N \sim \Nc({\bf 0}, \sigma_{T}^{2} \mathbf{I})$
\For{$i = N-1:0$} \do \\
\State $\xb_i \gets \text{Predictor}(\xb_{i+1}, \sigma_i, \sigma_{i+1})$
\State $\xb_i \gets \operatorname{Re}(\xb_i + A^*(y - A\xb_i))$
\For{$j = 1:M$} \do \\
\State $\xb_i \gets \text{Corrector}(\xb_{i}, \sigma_i, \epsilon_i)$
\State $\xb_i \gets \operatorname{Re}(\xb_i + A^*(y - A\xb_i))$
\EndFor
\EndFor
\State \textbf{return} $\xb_0$
\end{algorithmic}\label{alg:PC-POCS}
\end{algorithm}

\begin{algorithm}[!t]
\caption{Score-based sampling (SENSE-type)}
\begin{algorithmic}[1]
\Require $\sb_{\theta}, N, M, \{\epsilon_i\}$ \Comment{step size}, $\{\sigma_i\}$ \Comment{noise schedule}
\If{\text{parallel imaging (PI)}} 
    \State $A := \Pc_{\Omega} \Fc \Sc$
\Else
    \State $A := \Pc_{\Omega} \Fc$
\EndIf
\State $\xb_N \sim \Nc({\bf 0}, \sigma_{T}^{2} \mathbf{I})$
\For{$i = N-1:0$} \do \\
\State $\operatorname{Re}(\xb_i) \gets \text{Predictor}(\operatorname{Re}(\xb_{i+1}), \sigma_i, \sigma_{i+1})$
\State $\operatorname{Im}(\xb_i) \gets \text{Predictor}(\operatorname{Im}(\xb_{i+1}), \sigma_i, \sigma_{i+1})$
\State $\xb_i = \operatorname{Re}(\xb_i) + \iota \operatorname{Im}(\xb_i)$
\State $\xb_i \gets \xb_i + A^*(y - A\xb_i)$
\For{$j = 1:M$} \do \\
\State $\operatorname{Re}(\xb_i) \gets \text{Corrector}(\operatorname{Re}(\xb_{i}), \sigma_i, \epsilon_i)$
\State $\operatorname{Im}(\xb_i) \gets \text{Corrector}(\operatorname{Im}(\xb_{i}), \sigma_i, \epsilon_i)$
\State $\xb_i = \operatorname{Re}(\xb_i) + \iota \operatorname{Im}(\xb_i)$
\State $\xb_i \gets \xb_i + A^*(y - A\xb_i)$
\EndFor
\EndFor
\State \textbf{return} $\xb_0$
\end{algorithmic}\label{alg:PC-POCS_complex}
\end{algorithm}

Unfortunately, this algorithm can only be used when we know a priori that the signal only contains real values, and care must be taken since in most practical situations of MRI reconstruction, the signal that we would like to reconstruct is complex. This introduces a caveat when reconstructing the data with the score function because the original theory of score-based SDEs~\citep{song2020score} did not consider complex signals.

One approach that is feasible is to train a score function so that it handles complex signals, was proposed in \citep{ramzi2020denoising}. Implementation-wise, this corresponds to considering real and imaginary parts of the signal as separate channels and applying the de-noising score matching objective to handle $2 \times H \times W$ sized image, where $H$ and $W$ are the height, and the width of the image, respectively. However, we empirically found that this treatment reduces the stability of network training, and also hurts the performance of the reconstruction using Algorithm \ref{alg:PC-POCS}. We further note that this treatment limits the practicality since the model now requires raw $k$-space data for training.

To overcome these limitations, we propose a simple fix to Algorithm \ref{alg:PC-POCS}, which provides a way to use the score function $\sb_\thetab$ trained with magnitude images {\em only}, and use it to reconstruct complex images. The method is presented in Algorithm \ref{alg:PC-POCS_complex}, where we split the image into real and imaginary parts, and apply the predictor-corrector step separately to each part. Accordingly, we can use the same score function that was trained with magnitude images to deal with complex image data in a seamless way. This simple fix works surprisingly well, and we show in Section~\ref{sec:result_single} that reconstruction of complex-valued coil data with Algorithm \ref{alg:PC-POCS_complex} even outperforms the standard feed-forward neural network trained with explicit supervision. Being able to utilize score functions trained with magnitude-only data to reconstruct complex-valued data is a great advantage since we can use only the DICOM data to train the neural network. This is advantageous because the plethora of MR scans exist in the form of DICOM~\citep{zbontar2018fastmri}, while the raw $k$-space data are usually discarded due to their excessive memory size.

\subsection{Diffusion model meets Parallel Imaging (PI)}

\begin{algorithm}[!t]
\caption{Score-based sampling (SSOS-type)}
\begin{algorithmic}[1]
\Require $\sb_{\theta}, N, \{\epsilon_i\}$ \Comment{step size}, $\{\sigma_i\}$ \Comment{noise schedule}\\
\textbf{Define} $A := \Pc_{\Omega} \Fc$ 
\State $\xb^{(k)}_N \sim \Nc({\bf 0}, \sigma_{T}^{2} \mathbf{I})$
\For{$i = N-1:0$} \do \\
\For{$k = 1:c$} \do (\textbf{parallel})
\State $\operatorname{Re}(\xb^{(k)}_i) \gets \text{Predictor}(\operatorname{Re}(\xb^{(k)}_{i+1}), \sigma_i, \sigma_{i+1})$
\State $\operatorname{Im}(\xb^{(k)}_i) \gets \text{Predictor}(\operatorname{Im}(\xb^{(k)}_{i+1}), \sigma_i, \sigma_{i+1})$
\State $\xb^{(k)}_i = \operatorname{Re}(\xb^{(k)}_i) + \iota \operatorname{Im}(\xb^{(k)}_i)$
\State $\xb^{(k)}_i \gets \xb_i^{(k) + A^*(y^{(k)} - A\xb_i^{(k)})}$
\State $\operatorname{Re}(\xb^{(k)}_i) \gets \text{Corrector}(\operatorname{Re}(\xb^{(k)}_{i}), \sigma_i, \epsilon_i)$
\State $\operatorname{Im}(\xb^{(k)}_i) \gets \text{Corrector}(\operatorname{Im}(\xb^{(k)}_{i}), \sigma_i, \epsilon_i)$
\State $\xb^{(k)}_i = \operatorname{Re}(\xb^{(k)}_i) + \iota \operatorname{Im}(\xb^{(k)}_i)$
\State $\xb^{(k)}_i \gets \xb_i^{(k)} + A^*(y^{(k)} - A\xb_i^{(k)})$
\EndFor
\EndFor
\State $\xb_0 = \sqrt{\sum_{k=1}^c |\xb^{(c)}_0|^2}$ \Comment{SSOS}
\State \textbf{return} $\xb_0$
\end{algorithmic}\label{alg:PC-POCS_GRAPPA}
\end{algorithm}

\begin{algorithm}[!t]
\caption{Score-based sampling (Hybrid-type)}
\begin{algorithmic}[1]
\Require $\sb_{\theta}, N, M, m, \{\epsilon_i\}$ \Comment{step size}, $\{\sigma_i\}$ \Comment{noise schedule}
\textbf{Define} $A_{(sc)} := \Pc_{\Omega} \Fc$ \\
\textbf{Define} $A_{(mc)} := \Pc_{\Omega} \Fc \Sc$ 
\State $\xb^{(k)}_N \sim \Nc({\bf 0}, \sigma_{T}^{2} \mathbf{I})$
\For{$i = N-1:0$} \do \\
\For{$k = 1:c$} \do (\textbf{parallel})
\State $\operatorname{Re}(\xb^{(k)}_i) \gets \text{Predictor}(\operatorname{Re}(\xb^{(k)}_{i+1}), \sigma_i, \sigma_{i+1})$
\State $\operatorname{Im}(\xb^{(k)}_i) \gets \text{Predictor}(\operatorname{Im}(\xb^{(k)}_{i+1}), \sigma_i, \sigma_{i+1})$
\State $\xb^{(k)}_i = \operatorname{Re}(\xb^{(k)}_i) + \iota \operatorname{Im}(\xb^{(k)}_i)$
\State $\xb^{(k)}_i \gets \xb_i^{(k) + A_{(sc)}^{*}(y^{(k)} - A_{(sc)}\xb_i^{(k)})}$
\State $\operatorname{Re}(\xb^{(k)}_i) \gets \text{Corrector}(\operatorname{Re}(\xb^{(k)}_{i}), \sigma_i, \epsilon_i)$
\State $\operatorname{Im}(\xb^{(k)}_i) \gets \text{Corrector}(\operatorname{Im}(\xb^{(k)}_{i}), \sigma_i, \epsilon_i)$
\State $\xb^{(k)}_i = \operatorname{Re}(\xb^{(k)}_i) + \iota \operatorname{Im}(\xb^{(k)}_i)$
\State $\xb^{(k)}_i \gets \xb_i^{(k)} + A_{(sc)}^{*}(y^{(k)} - A_{(sc)}\xb_i^{(k)})$
\EndFor
\If{$\mathrm{mod}(i, m) == 0$} \do \\
\State $\xb_i = [\xb^{(1)}, \xb^{(2)}, \dots, \xb^{(c)}]$ \Comment{Aggregation}
\State $\xb_i \gets \xb_i + \lambda A_{(mc)}^*(y - A_{(mc)}\xb_i)$
\EndIf
\EndFor
\State $\xb_0 = \sqrt{\sum_{k=1}^c |\xb^{(c)}_0|^2}$ \Comment{SSOS}
\State \textbf{return} $\xb_0$
\end{algorithmic}\label{alg:PC-POCS_hybrid}
\end{algorithm}

\begin{figure}[!hbt]
    \centering\includegraphics[width=9cm]{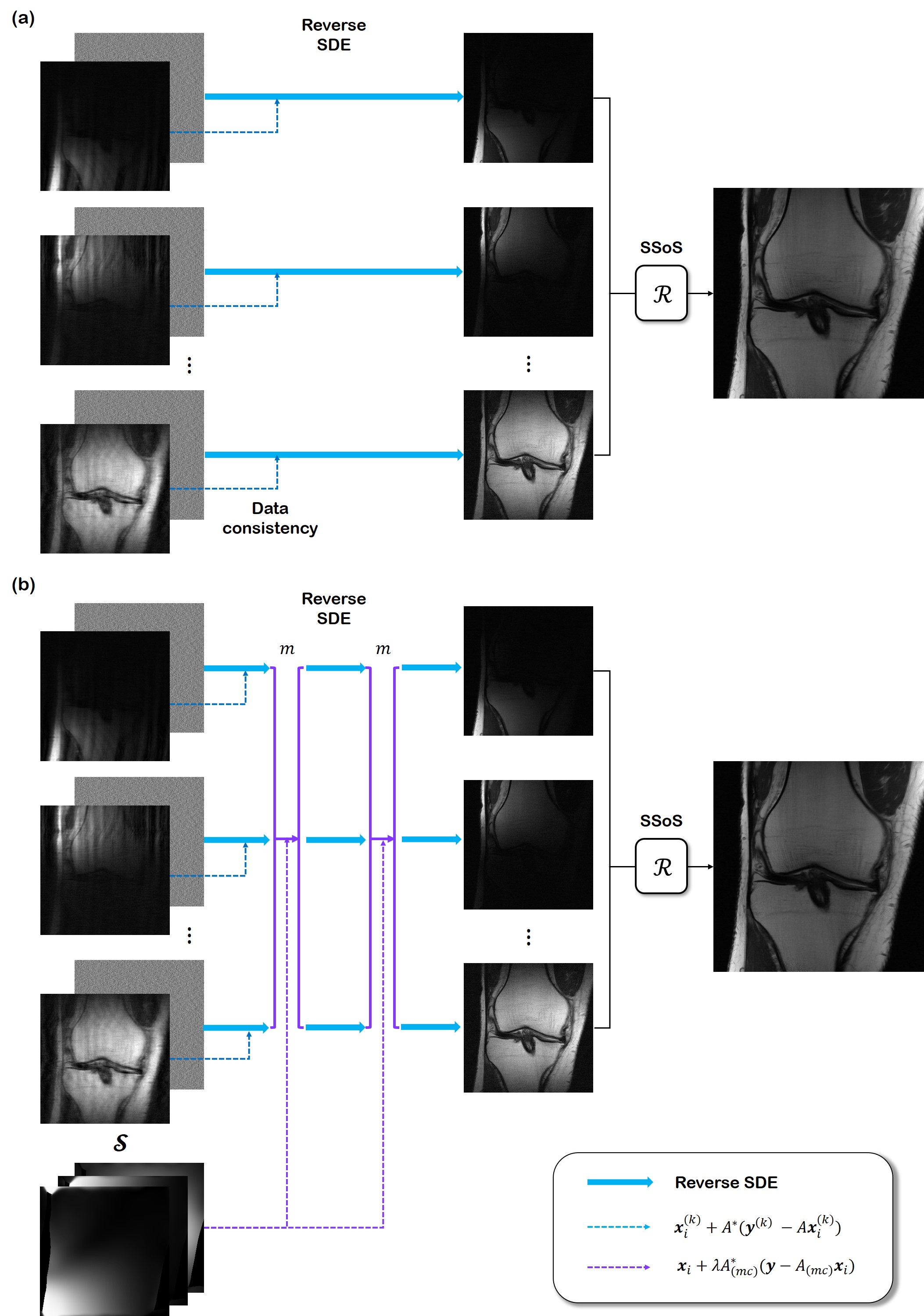}
    \caption{Illustration of parallel imaging applications. (a) SSOS-type sampling. Coil images are reconstructed separately, then merged with the SSOS operation. (b) Hybrid-type sampling. Dependency between the coil images is injected every $m$ steps of iteration.}
	\label{fig:score_PI}
\end{figure}

While the proposed score-based framework for the reconstruction of complex-valued data is extremely useful, most modern MRI scanners~\citep{zbontar2018fastmri} have {\em multiple} receiver coils, which capture the signal with different sensitivities. Since the birth of PI~\citep{deshmane2012parallel}, myriad of techniques to reconstruct the true latent signal have been proposed in the literature, two of the most prominent being SENSE~\citep{pruessmann1999sense}, and GRAPPA~\citep{griswold2002generalized}.
The former approach requires estimation or calibration of coil sensitivity maps, which are incorporated in the forward model specified in \eqref{eq:forward_model}. Contrarily, the latter approach drops the need for the sensitivity maps, by simply taking the sum-of-root-sum-of-squares (SSOS) of each reconstructed coil image~\citep{roemer1990nmr}. This approach is one of the most widely used methods in clinical practice due to several advantages including implementation ease. Here, we show that our score-based method can be integrated seamlessly into the SSOS-type approach.

Specifically, as introduced in Algorithm~\ref{alg:PC-POCS_GRAPPA}, our objective is to reconstruct the data coil-by-coil.
 For PI, we set the number corrector step
to one for brevity, i.e. $M = 1$.
More specifically, unlike GRAPPA,  instead of estimating of the GRAPPA kernels, we simply apply Algorithm~\ref{alg:PC-POCS_complex} separately to each coil image, as illustrated in Fig.~\ref{fig:score_PI} (a). 
Notably, although our score function estimator $\sb_\thetab(\xb, t) \simeq \nabla_{\xb_t} \log p(\xb)$ were not trained with separate coil images, since the distribution of independent coil images does not deviate much from $p(\xb)$, simply applying Algorithm~\ref{alg:PC-POCS_complex} to each coil image induces a very accurate reconstruction.

One downside of Algorithm~\ref{alg:PC-POCS_GRAPPA} is that there exists no cross-talk between the coil reconstructions. This may be suboptimal, because the reconstruction process does not take into account that all the coil images come from the same image. 
Instead, in order to properly leverage the correlation between different coil images, we additionally implement a hybrid-type method in Algorithm~\ref{alg:PC-POCS_hybrid} by incorporating the SENSE type constraint. Specifically, for every $m$ steps of individual coil updates, we coerce dependency between the coils with the following update:
\begin{equation}\label{eq:gd}
    \xb_i \gets \xb_i + \lambda A_{(mc)}^*(y - A_{(mc)}\xb_i),
\end{equation}
where $\lambda \in [0, 1]$ decides the extent to which data consistency is imposed, and $A_{(mc)}$ is the standard  multi-coil forward measurement matrix as defined in~\eqref{eq:mm_mri}.  Once this is done, we arrive at the final result with SSOS operation. See Fig.~\ref{fig:score_PI} (b) for an illustration.
By applying update steps of \eqref{eq:gd} in between with linearly decreasing the value of $\lambda$ as the iteration progresses, we observe improved performance and sharper reconstructions.

Among the different sampling patterns, we see that Algorithm~\ref{alg:PC-POCS_hybrid} generally performs better with 1D sampling patterns, whereas Algorithm~\ref{alg:PC-POCS_GRAPPA} performs better with 2D patterns. Hence, we report on reconstructions with Algorithm~\ref{alg:PC-POCS_hybrid} for 1D under-sampling, and Algorithm~ \ref{alg:PC-POCS_GRAPPA} for 2D under-sampling.
One caveat for the SSOS-type, and hybrid-type approaches are the slow inference speed. Naively implementing the algorithms will induce $c$ times longer computation time compared to the single-coil reconstruction.
However, this can be much relieved by performing parallel computation with each coil data, since no computation needs to be shared across the threads. Given sufficient GPU resources, we expect that the computation time needed for PI reconstruction will be reduced down to the time needed for single-coil reconstruction.

\section{Methods}
\label{sec:methods}

\subsection{Experimental data}

The main experiments, including the training of the score function, were performed with fastMRI knee dataset~\citep{zbontar2018fastmri}, which is publicly available\footnote{\href{https://fastmri.org/}{https://fastmri.org/}}. We trained the network with 320 $\times$ 320 size target image magnitude, given as the key \code{reconstruction\_esc}. We note that it is possible to train the score function with the same target from \code{reconstruction\_rss} of the multi-coil dataset, but we found no significant difference in the performance.

Among 973 volumes of training data, we dropped the first and last five slices from each volume, to avoid training the model with noise-only data. This results in approximately 25k slices of training data. For testing, we randomly sampled 30 volumes from the validation set, and dropped the first and last 5 slices from each volume. For PI experiments, we resorted to 10 volumes, due to the computational limitations.

\subsection{Implementation details}
\label{sec:imple_details}

We follow similar procedures to train VE-SDE as advised in \citep{song2020score}. Specifically, we train the network with the objective given in Eq.~\eqref{eq:score_cost}, with setting $\lambda(t) = \sigma^2(t)$. Note that this specific choice of $\lambda(t)$ stabilizes the noise scale across $t$, and theoretically corresponds to likelihood weighting, as proven in~\citep{song2021maximum}. Plugging in the weighting function, we can train the model with the following cost:
\begin{align}\label{eq:score_cost_VESDE}
    \min_{\thetab} &\Ed_{t \sim U(\epsilon, 1)} \Ed_{\xb(0) \sim p_0} \Ed_{\xb(t) \sim \Nc(\xb(0),\sigma^2(t)\Ib)}\Big[ \\ \notag
    &\norm{\sigma(t)\sb_{\thetab} (\xb(t), t) - \frac{\xb(t) - \xb(0)}{\sigma(t)}}_2^2
    \Big],
\end{align}
with setting $\epsilon=10^{-5}$ to circumvent numerical issues.

For the step size $\epsilon_i$ used in the Langevin MC corrector step, we follow what is advised in~\citep{song2020score}, and set
\begin{equation}
    \epsilon_i = 2r\frac{\|\zb\|_2}{\|s_{\theta}(\xb_i, \sigma_i)\|_2},
\end{equation}
where $r = 0.16$ is set to a constant value.
For noise variance schedule, we fix $\sigma_{min} = 0.01, \sigma_{max} = 378$, which is similar to what is advised in the technique of~\citep{song2020improved}, then take the geometric series with the following form:
\begin{equation}
    \sigma(t) = \sigma_{min} \left(\frac{\sigma_{max}}{\sigma_{min}}\right)^t.
\end{equation}

We take the batch size of 1, and optimize the network using the \code{Adam} optimizer ($\beta_1 = 0.9, \beta_2 = 0.999$). We use a linear warm-up schedule for the parameters for the first 5000 steps of optimization, reaching \code{2e-4} at the 5000$^{\text{th}}$ step. We apply gradient clipping with the maximum value of $1.0$~\citep{arjovsky2017wasserstein}. Exponential moving average with rate 0.999 is applied to the parameters. Optimization was performed for 100 epochs, and it took about 3 weeks of training the score function with a single RTX 3090 GPU. All code was implemented in PyTorch~\citep{paszke2019pytorch}.

For all algorithms, we use $N = 2000, M = 1$ iterations for inference as default, unless specified otherwise. For the hybrid-type Algorithm~\ref{alg:PC-POCS_hybrid}, we start with $\lambda = 1.0$ in the first iteration, and linearly decrease the value to $\lambda = 0.2$ at the last iteration. Single forward-pass through $\sb_\thetab$ is required for both predictor and corrector steps, which takes about 150 ms with a commodity GPU. Summing up, this results in about 10 minutes of reconstruction time for real-valued images, and 20 minutes of reconstruction time for complex-valued images. We discuss ways for speeding up inference, and some potential directions for future studies in Section~\ref{sec:speed}.

\subsection{Model Architecture}

We base the implementation of the time-dependent score function model \code{ncsnpp} \footnote{\href{https://github.com/yang-song/score_sde_pytorch}{https://github.com/yang-song/score\_sde\_pytorch}} as suggested in~\citep{song2020score}. The model architecture stems from U-Net~\citep{ronneberger2015u}, and the sub-block which consist U-Net are adopted from residual blocks of BigGAN~\citep{brock2018large}. The skip connections in the residual blocks are scaled by $1 / \sqrt{2}$ as in~\citep{karras2017progressive, karras2019style, karras2020analyzing}. For pooling and unpooling, we adopt anti-aliasing pooling of~\citep{zhang2019making}. The resulting U-Net has 4 different levels of scale, with 4 residual networks at each level. Conditioning of network with the time index $t$ is performed with Fourier features~\citep{tancik2020fourier}, where the conditional features are added to the encoder features. For further details, see Appendix.

\subsection{Comparison study}

To verify superiority over the current standards, we perform comparison studies with baseline methods used in~\citep{zbontar2018fastmri}. We choose total variation (TV) reconstruction~\citep{block2007undersampled} as the representative CS reconstruction method, where we use the implementation in \code{sigpy.mri.app.TotalVariationRecon}\footnote{\href{https://github.com/mikgroup/sigpy}{https://github.com/mikgroup/sigpy}}. We perform grid search on the hyper-parameter \code{lambda}, and report only the best results among them.

For a representative deep learning approach, we use supervised learning-based reconstruction using U-Net~\citep{zbontar2018fastmri}. While we could use the open-sourced\footnote{\href{https://github.com/facebookresearch/fastMRI}{https://github.com/facebookresearch/fastMRI}} pre-trained model, we re-implemented the model to achieve better performance.

We additionally compare against the state-of-the-art supervised methods. For the real-valued simulation study and single-coil experiment, we compare against DuDoRNet~\citep{zhou2020dudornet}. We use the official implementation\footnote{\href{https://github.com/bbbbbbzhou/DuDoRNet}{https://github.com/bbbbbbzhou/DuDoRNet}}, with 4 recurrent blocks and default parameters. We resort to the same proton density (PD)/ proton density fat suppressed (PDFS) image for prior information. For state-of-the-art parallel imaging method, we use end-to-end variational network (E2E-varnet)~\citep{sriram2020end}, which simultaneously estimates the sensitivity maps. We use the official fastMRI github with default parameters as advised with the fastMRI knee dataset. For all the deep learning comparison studies, we train the network with Gaussian 1D random sampling masks.

\subsection{Measurement of Reconstruction Quality}

To quantify the proximity of the reconstructions to the target, we use the standard metrics - peak signal-to-noise ratio (PSNR), and structural similarity index (SSIM). We further test the statistical significance of the differences using repeated measures analysis of variance (RM-ANOVA) with the MedCalc software~\citep{schoonjans1995medcalc}.
While these are the two most widely used metrics in the community, it is also well known that these metrics hardly line up with the radiologists' scoring on the image quality~\cite{mason2019comparison}. 

To fully capture the superiority of the proposed method, we focus on the fact that \textit{good} reconstructions are the ones that can be used for \textit{accurate diagnosis}. 
If the reconstruction quality closely matches the ground truth, there should be no degeneration in the performance of the downstream tasks - in our case diagnosis.
In order to compare against ground truth, the diagnostic ability of each reconstruction, we leverage the recent fastMRI+~\citep{zhao2021fastmri+} dataset. For fastMRI knee data, fastMRI+ annotations provide bounding boxes around the pathologic region. We train a standard object detection model using the ground truth (fully sampled) images and use this network to compare how well the model performs on pathology detection with reconstructions using different methods. By doing so, we can measure the amount of distribution shift that occurs with each reconstruction method. The less the difference in the performance, we conclude that there is less distribution shift from the fully-sampled data. For detailed experimental procedures, see the following section, i.e. Section \ref{sec:yolo}. From the experiments, we quantify three standard metrics from the object detection literature - mean average precision (mAP), precision, and recall.

\subsection{Pathology Detection}
\label{sec:yolo}

For the object detection model, we use the state-of-the-art YOLO v5\footnote{\href{https://github.com/ultralytics/yolov5}{https://github.com/ultralytics/yolov5}}. We use the default configuration of \code{YOLOv5m}, which is a medium-sized model, often suggested as the baseline model when you do not have a sufficient amount of data. When trying to fit larger versions of the model, namely \code{YOLOv5l}, \code{YOLOv5x}, etc., we found that overfitting occurs and the performance drops by a small margin. 

For training data of the YOLOv5 model, we use all the training data with annotations in fastMRI+, which consists of 8053 images in total. We do not include any images without annotations in the training set. For testing, we select 15 random cases out of the validation set.

Model weights were fine-tuned from the open-source pre-trained model, and were trained for 300 epochs using the batch size of 16. Training took about a day on $2\times$2080Ti GPU.

\subsection{Generalization to different anatomy and contrast}
\label{sec:va}

One observation that can be made is that the proposed method, which utilizes the score function as the main workhorse of the algorithm, is robust to distribution shifts. Otherwise, Algorithms~\ref{alg:PC-POCS_complex},\ref{alg:PC-POCS_GRAPPA},\ref{alg:PC-POCS_hybrid} would not have worked in the first place, since the training data distribution and the inference data distribution are different. Subsequently, one might wonder how far we are able to push this discrepancy and still achieve satisfactory results. To further investigate the generalization capability of the proposed method, we ran extensive experiments with data collected from different anatomy and contrast.
To achieve maximal diversity, we collected data from various open-source database, including mridata~\citep{mridata}, human connectome project (HCP) MRI dataset (\href{http://db.humanconnectome.org}{http://db.humanconnectome.org}, \href{http://github.com/hkaggarwal/modl}{http://github.com/hkaggarwal/modl}, and MASSIVE\footnote{\textbf{M}ultiple \textbf{A}cquisition for \textbf{S}tandardization of \textbf{S}tructural \textbf{I}maging \textbf{V}alidation and \textbf{E}valuation} (\href{http://massive-data.org/index.html}{http://massive-data.org/index.html}). For experiments, all data were retrospectively down-sampled from the fully-sampled $k$-space.

\begin{figure*}[!hbt]
    \centering\includegraphics[width=18cm]{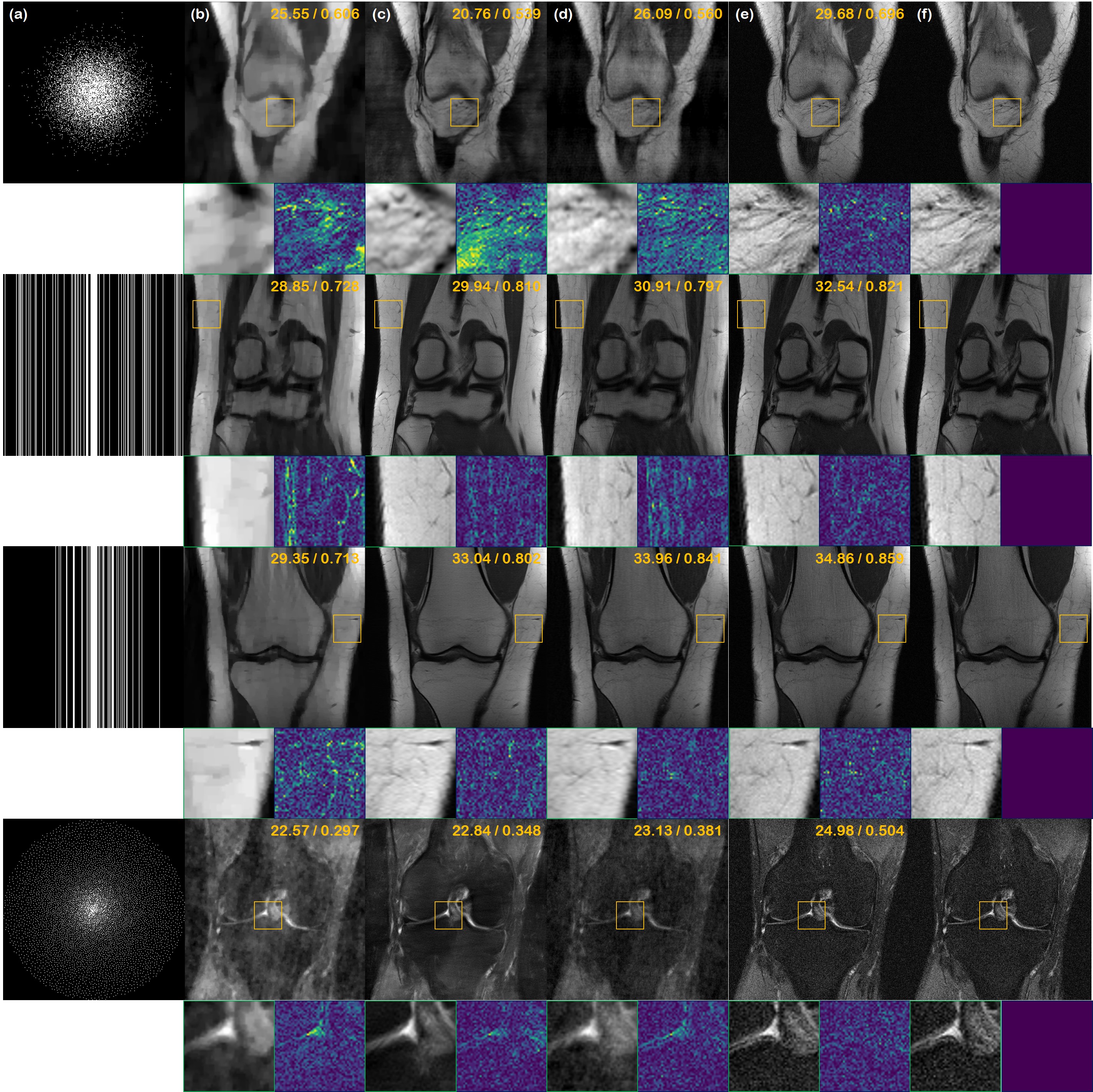}
    \caption{Reconstructions of the real-valued simulation study. (a) Sub-sampling mask used to generate under-sampled image, (b) TV, (c) supervised learning (U-Net) (d) DuDoRNet~\citep{zhou2020dudornet}, (e) proposed method,  and (f) ground truth. 1$^\text{st}$ row: 2D $\times 8$ Gaussian random sampling, 2$^\text{nd}$ row: 1D $\times 4$ uniform random sampling, 3$^\text{rd}$ row: 1D $\times 8$ Gaussian random sampling, 4$^\text{th}$ row: $\times 15$ variable density poisson disk sampling. Green box: Zoom in version of the indicated yellow box, Blue box: Difference magnitude of the inset (in \code{Viridis} colormap). Yellow numbers in the upper right corner indicate PSNR [db], and SSIM, respectively.}
	\label{fig:real_results}
\end{figure*}

\begin{figure*}[!hbt]
    \centering\includegraphics[width=18cm]{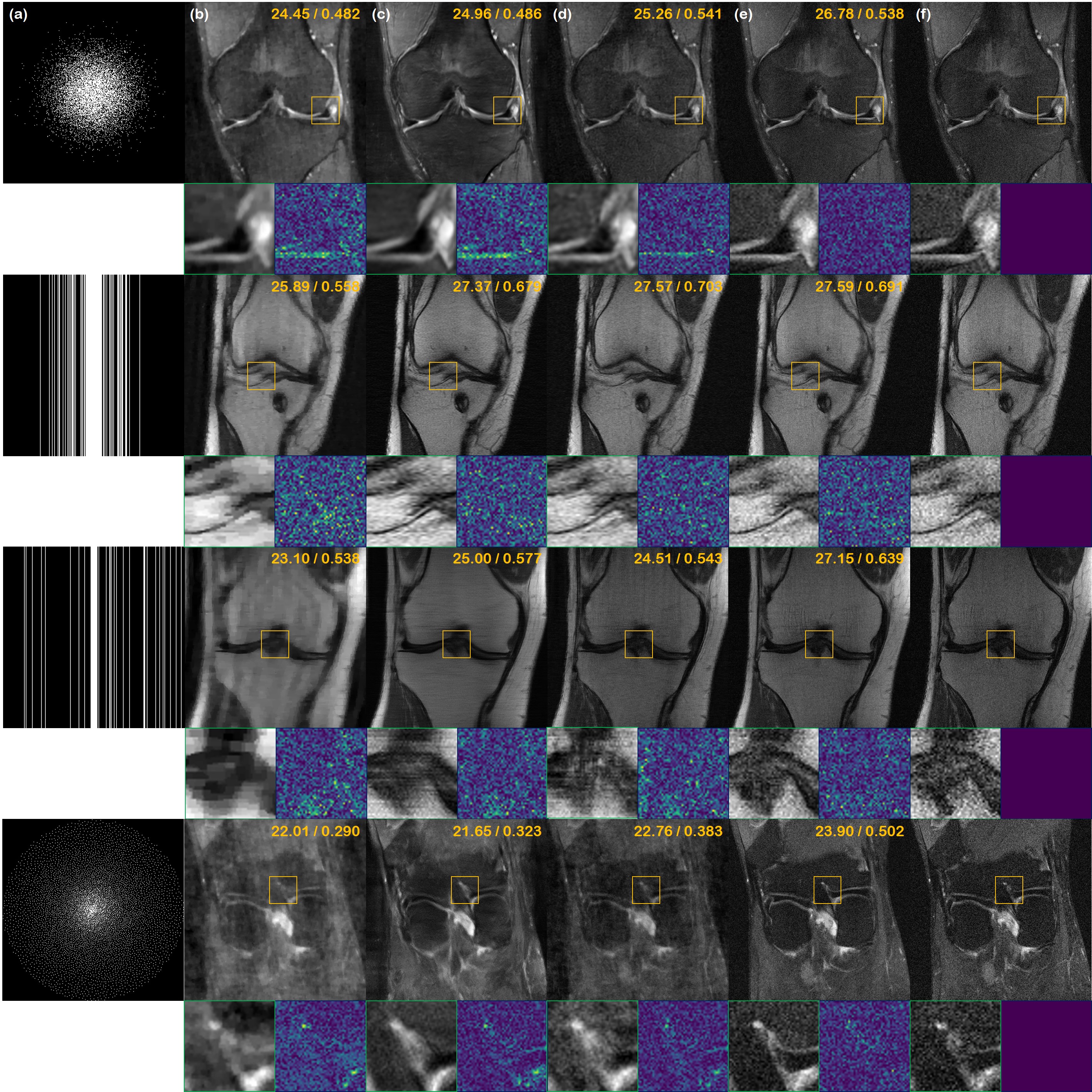}
    \caption{Single-coil complex-valued image reconstruction results. (a) Sub-sampling mask used to generate under-sampled image, (b) TV, (c) supervised learning (U-Net) (d) DuDoRNet~\citep{zhou2020dudornet}, (e) proposed method,  and (f) ground truth. 1$^\text{st}$ row: 2D $\times 8$ Gaussian random sampling, 2$^\text{nd}$ row: 1D $\times 4$ Gaussian random sampling, 3$^\text{rd}$ row: 1D $\times 8$ uniform random sampling, 4$^\text{th}$ row: $\times 15$ variable density poisson disk sampling. Green box: Zoom in version of the indicated yellow box, Blue box: Difference magnitude of the inset (in \code{Viridis} colormap). Yellow numbers in the upper right corner indicate PSNR [db], and SSIM, respectively.}
	\label{fig:singlecoil_results}
\end{figure*}

\section{Results}
\label{sec:results}

\begin{table*}[!hbt]
\centering
\begin{adjustbox}{max width=0.95\textwidth}
\begin{tabular}{c|c|c|c|cccc|c}
                             & &                      &               & TV & Supervised U-Net & DuDoRNet & Proposed & $p^{*} < 0.0001$ \\ \hline
                             \multirow{16}{*}{Simulation (real)} &
\multirow{4}{*}{Uniform 1D}  & \multirow{2}{*}{$\times$ 4} & PSNR {[}db{]} &  27.55 $\pm$ 1.60  &  29.21 $\pm$ 2.09  & 30.79 $\pm$ 0.91  & \textbf{31.10} $\pm$ 1.60 & \multirow{2}{*}{TV, U-Net}         \\
                             & &                      & SSIM          & 0.667 $\pm$ 0.045   &  0.788 $\pm$ 0.030  &   \textbf{0.796} $\pm$ 0.022    &  0.795 $\pm$ 0.036 &         \\ \cline{3-9}&&&&&&&&\\[-1em]
                             & & \multirow{2}{*}{$\times$ 8} & PSNR {[}db{]} &  26.35 $\pm$ 1.74  &  27.40 $\pm$ 2.39  &  24.08 $\pm$ 1.80      &    \textbf{28.37} $\pm$ 1.97  & \multirow{2}{*}{TV, U-Net, DuDoRNet}    \\
                             & &                      & SSIM          & 0.631 $\pm$ 0.052   & 0.722 $\pm$ 0.042   &  0.603 $\pm$ 0.037     &  \textbf{0.771} $\pm$ 0.051&         \\
                             \cline{2-9}&&&&&&&&\\[-1em]
& \multirow{4}{*}{Gaussian 1D} & \multirow{2}{*}{$\times$ 4} & PSNR {[}db{]} & 30.77 $\pm$ 1.24   & 32.85 $\pm$ 1.25  &  33.01 $\pm$ 1.05&  \textbf{33.32} $\pm$ 1.22 & \multirow{2}{*}{TV, U-Net, DuDoRNet}       \\
                             & &                      & SSIM          & 0.752 $\pm$ 0.029   & 0.829 $\pm$ 0.023   & \textbf{0.858} $\pm$ 0.016      &  0.825 $\pm$ 0.028    &     \\
                             \cline{3-9}&&&&&&&&\\[-1em]
                             & & \multirow{2}{*}{$\times$ 8} & PSNR {[}db{]} &  28.87 $\pm$ 1.56  &  30.81 $\pm$ 1.37  & 30.46 $\pm$ 1.28      &     \textbf{30.94} $\pm$ 1.17   & \multirow{2}{*}{TV, DuDoRNet}   \\
                             & &                      & SSIM          &  0.70 $\pm$ 0.041  & \textbf{0.779} $\pm$ 0.029  &  0.776 $\pm$ 0.025    &    0.761 $\pm$ 0.025 &      \\
                             \cline{2-9}&&&&&&&&\\[-1em]
& \multirow{4}{*}{Gaussian 2D} & \multirow{2}{*}{$\times$ 8} & PSNR {[}db{]} & 23.19 $\pm$ 2.30    & 21.92 $\pm$ 4.51   & 25.29 $\pm$ 4.16      &     \textbf{29.95} $\pm$ 2.04     & \multirow{2}{*}{TV, U-Net, DuDoRNet}\\
                             & &                      & SSIM          &  0.599 $\pm$ 0.082  & 0.573 $\pm$ 0.173   &  0.643 $\pm$ 0.077     &  \textbf{0.701} $\pm$ 0.071        & \\
                             \cline{3-9}&&&&&&&&\\[-1em]
                             & & \multirow{2}{*}{$\times$ 15} & PSNR {[}db{]} &  18.20 $\pm$ 2.80  & 17.43 $\pm$ 0.45  &  21.24 $\pm$ 2.09 &  \textbf{29.58} $\pm$ 1.47     & \multirow{2}{*}{TV, U-Net, DuDoRNet}   \\
                             & &                      & SSIM          &  0.456 $\pm$ 0.096  &  0.451 $\pm$ 0.158  &  0.518 $\pm$ 0.046     &         \textbf{0.678} $\pm$ 0.052 & \\ \cline{2-9}&&&&&&&&\\[-1em]
& \multirow{4}{*}{VD Poisson disk} & \multirow{2}{*}{$\times$ 8} & PSNR {[}db{]} & 21.26 $\pm$ 0.96    & 23.47 $\pm$ 1.07   & 23.22 $\pm$ 0.93      &     \textbf{31.83} $\pm$ 1.15 & \multirow{2}{*}{TV, U-Net, DuDoRNet}    \\
                             & &                      & SSIM          &  0.516 $\pm$ 0.038  & 0.606 $\pm$ 0.083   &  0.516 $\pm$ 0.063     &  \textbf{0.769} $\pm$ 0.030    &     \\
                             \cline{3-9}&&&&&&&&\\[-1em]
                             & & \multirow{2}{*}{$\times$ 15} & PSNR {[}db{]} &  19.56 $\pm$ 0.51  & 20.97 $\pm$ 1.73  &  22.84 $\pm$ 2.34 &  \textbf{30.46} $\pm$ 1.21& \multirow{2}{*}{TV, U-Net, DuDoRNet}        \\
                             & &                      & SSIM          &  0.517 $\pm$ 0.042  &  0.620 $\pm$ 0.054  &  0.561 $\pm$ 0.063     &         \textbf{0.709} $\pm$ 0.035&  \\ \hline
                             
\multirow{16}{*}{Single-coil} & \multirow{4}{*}{Uniform1D} & \multirow{2}{*}{$\times$ 4} & {PSNR {[}db{]}} & 27.13 $\pm$ 1.91 & 30.90 $\pm$ 1.78 & 30.43 $\pm$ 0.79 & \textbf{31.95} $\pm$ 1.45 & \multirow{2}{*}{TV, U-Net, DuDoRNet} \\
& & & SSIM & 0.636 $\pm$ 0.041 & 0.801 $\pm$ 0.027 & 0.793 $\pm$ 0.024 & \textbf{0.812} $\pm$ 0.036 &  \\ \cline{3-9}&&&&&&&&\\[-1em]
& & \multirow{2}{*}{$\times 8$} & PSNR {[}db{]} & 24.38 $\pm$ 2.01 & 27.48 $\pm$ 1.66 & 24.72 $\pm$ 1.89 & \textbf{27.97} $\pm$ 2.03 & \multirow{2}{*}{TV, DuDoRNet} \\
& & & SSIM & 0.601 $\pm$ 0.090 & 0.720 $\pm$ 0.039 & 0.641 $\pm$ 0.055 & \textbf{0.738} $\pm$ 0.053 &  \\ \cline{2-9}&&&&&&&&\\[-1em]
& \multirow{4}{*}{Gaussian 1D} & \multirow{2}{*}{$\times$ 4} & PSNR {[}db{]} & 28.39 $\pm$ 1.09 & 32.86 $\pm$ 1.23 & 33.46 $\pm$ 1.35 & \textbf{33.96} $\pm$ 1.27 & \multirow{2}{*}{TV, U-Net, DuDoRNet} \\
& & & SSIM & 0.679 $\pm$ 0.077 & 0.828 $\pm$ 0.024 & \textbf{0.856} $\pm$ 0.022 & 0.858 $\pm$ 0.028 &  \\ \cline{3-9}&&&&&&&&\\[-1em]
& & \multirow{2}{*}{$\times$ 8} & PSNR {[}db{]} & 25.91 $\pm$ 3.21 & 30.80 $\pm$ 1.34 & 29.65 $\pm$ 1.76 & \textbf{30.82} $\pm$ 1.37 & \multirow{2}{*}{TV, DuDoRNet}\\
& & & SSIM & 0.622 $\pm$ 0.050 & \textbf{0.777} $\pm$ 0.028 & 0.777 $\pm$ 0.028 & 0.762 $\pm$ 0.034 & \\
\cline{2-9}&&&&&&&&\\[-1em]
& \multirow{4}{*}{Gaussian 2D} & \multirow{2}{*}{$\times$ 8} & PSNR {[}db{]} & 20.09 $\pm$ 6.13 & 19.99 $\pm$ 5.12 & 21.53 $\pm$ 10.86 & \textbf{29.45} $\pm$ 2.97 & \multirow{2}{*}{TV, U-Net, DuDoRNet}\\
& & & SSIM & 0.592 $\pm$ 0.140 & 0.520 $\pm$ 0.187 & 0.541 $\pm$ 0.152 & \textbf{0.676} $\pm$ 0.118 & \\
\cline{3-9}&&&&&&&&\\[-1em]
& & \multirow{2}{*}{$\times$ 15} & PSNR {[}db{]} & 17.99 $\pm$ 3.15 & 17.24 $\pm$ 4.01 & 18.86 $\pm$ 5.51 & \textbf{26.15 }$\pm$ 4.44 & \multirow{2}{*}{TV, U-Net, DuDoRNet}\\
& & & SSIM & 0.460 $\pm$ 0.250 & 0.436 $\pm$ 0.156 & 0.490 $\pm$ 0.192 & \textbf{0.587}$\pm$ 0.148 & \\
\cline{2-9}&&&&&&&&\\[-1em]
& \multirow{4}{*}{VD Poisson disk} & \multirow{2}{*}{$\times$ 8} & PSNR {[}db{]} & 20.90 $\pm$ 3.92    & 19.67 $\pm$ 1.72   & 22.91 $\pm$ 2.54      &     \textbf{31.50} $\pm$ 1.24 & \multirow{2}{*}{TV, U-Net, DuDoRNet}    \\
                             & &                      & SSIM          &  0.626 $\pm$ 0.101  & 0.604 $\pm$ 0.067   &  0.643 $\pm$ 0.041     &  \textbf{0.762} $\pm$ 0.030    &     \\
                             \cline{3-9}&&&&&&&&\\[-1em]
                             & & \multirow{2}{*}{$\times$ 15} & PSNR {[}db{]} &  22.13 $\pm$ 1.92  & 21.67 $\pm$ 1.06  &  23.95 $\pm$ 1.13 &  \textbf{30.66} $\pm$ 1.30 & \multirow{2}{*}{TV, U-Net, DuDoRNet}       \\
                             & &                      & SSIM          &  0.616 $\pm$ 0.089  &  0.573 $\pm$ 0.098  &  0.593 $\pm$ 0.032     &         \textbf{0.706} $\pm$ 0.035 &  \\
\hline

\end{tabular}
\end{adjustbox}
\caption{{Quantitative metrics of real-valued simulation study and single-coil complex-valued image experiment. Numbers are presented as mean value $\pm$ unit standard deviation. Numbers in bold face indicate the best metric out of all the methods. $p$-values indicate results from RM-ANOVA. Proposed method statistically relevant against methods listed in the last column in terms of PSNR.
$^*$: Bonferroni corrected.}}
\label{tbl:quantitative_metric_single}
\end{table*}

\begin{table*}[!hbt]
\centering
\begin{adjustbox}{max width=0.95\textwidth}
\begin{tabular}{c|c|c|c|cccc|c}
                             & &                      &               & TV & Supervised U-Net & E2E-Varnet & Proposed & $p^{*} < 0.0001$ \\ \hline
                             \multirow{16}{*}{Multi-coil} &
\multirow{4}{*}{Uniform 1D}  & \multirow{2}{*}{$\times$ 4} & PSNR {[}db{]} &  27.32 $\pm$ 1.30  &  31.77 $\pm$ 2.30  & 32.96 $\pm$ 1.46  & \textbf{33.25} $\pm$ 1.42         & \multirow{2}{*}{TV, U-Net}\\
                             & &                      & SSIM          & 0.662 $\pm$ 0.112   &  0.846 $\pm$ 0.031  &   0.856 $\pm$ 0.072    &  \textbf{0.857} $\pm$ 0.201    &     \\ \cline{3-9}&&&&&&&&\\[-1em]
                             & & \multirow{2}{*}{$\times$ 8} & PSNR {[}db{]} &  25.02 $\pm$ 2.90  &  29.51 $\pm$ 3.99  &  31.98 $\pm$ 1.59      &    \textbf{32.01} $\pm$ 1.55   & \multirow{2}{*}{TV, U-Net}   \\
                             & &                      & SSIM          & 0.532 $\pm$ 0.158   & 0.780 $\pm$ 0.058   &  \textbf{0.828} $\pm$ 0.051     &  0.821 $\pm$ 0.071   &      \\
                             \cline{2-9}&&&&&&&&\\[-1em]
& \multirow{4}{*}{Gaussian 1D} & \multirow{2}{*}{$\times$ 4} & PSNR {[}db{]} & 30.55 $\pm$ 1.75   & 32.66 $\pm$ 1.18  &  33.15 $\pm$ 2.92     &  \textbf{34.25} $\pm$ 1.74        & \multirow{2}{*}{TV, U-Net, E2E-Varnet}\\
                             & &                      & SSIM          & 0.789 $\pm$ 0.128   & 0.866 $\pm$ 0.020   & 0.878 $\pm$ 0.062      &  \textbf{0.885} $\pm$ 0.024    &     \\
                             \cline{3-9}&&&&&&&&\\[-1em]
                             & & \multirow{2}{*}{$\times$ 8} & PSNR {[}db{]} &  27.98 $\pm$ 2.12  &  31.64 $\pm$ 1.27  & \textbf{33.15} $\pm$ 2.64      &     32.43 $\pm$ 1.69 & \multirow{2}{*}{TV, U-Net}    \\
                             & &                      & SSIM          &  0.747 $\pm$ 0.114  & 0.841 $\pm$ 0.021  &  \textbf{0.868} $\pm$ 0.069    &    0.855 $\pm$ 0.021    &   \\
                             \cline{2-9}&&&&&&&&\\[-1em]
& \multirow{4}{*}{Gaussian 2D} & \multirow{2}{*}{$\times$ 8} & PSNR {[}db{]} & 29.20 $\pm$ 1.85    & 24.51 $\pm$ 3.16   & 20.97 $\pm$ 2.98      &     \textbf{31.43} $\pm$ 1.70 & \multirow{2}{*}{TV, U-Net, E2E-Varnet}    \\
                             & &                      & SSIM          &  0.781 $\pm$ 0.012  & 0.724 $\pm$ 0.103   &  0.642 $\pm$ 0.085     &  \textbf{0.831} $\pm$ 0.036    &     \\
                             \cline{3-9}&&&&&&&&\\[-1em]
                             & & \multirow{2}{*}{$\times$ 15} & PSNR {[}db{]} &  26.28 $\pm$ 1.02  & 14.93 $\pm$ 0.95  &  16.66 $\pm$ 3.01 &  \textbf{29.17} $\pm$ 2.19     & \multirow{2}{*}{TV, U-Net, E2E-Varnet}   \\
                             & &                      & SSIM          &  0.547 $\pm$ 0.072  &  0.372 $\pm$ 0.070  &  0.435 $\pm$ 0.106     &         \textbf{0.704} $\pm$ 0.042 & \\ \cline{2-9}&&&&&&&&\\[-1em]
& \multirow{4}{*}{VD Poisson disk} & \multirow{2}{*}{$\times$ 8} & PSNR {[}db{]} & 29.52 $\pm$ 1.39    & 20.89 $\pm$ 1.10   & 20.70 $\pm$ 1.20      &     \textbf{31.98} $\pm$ 2.07 & \multirow{2}{*}{TV, U-Net, E2E-Varnet}    \\
                             & &                      & SSIM          &  0.562 $\pm$ 0.078  & 0.576 $\pm$ 0.063   &  0.592 $\pm$ 0.045     &  \textbf{0.816} $\pm$ 0.022 &         \\
                             \cline{3-9}&&&&&&&&\\[-1em]
                             & & \multirow{2}{*}{$\times$ 15} & PSNR {[}db{]} &  26.19 $\pm$ 1.19  & 16.01 $\pm$ 1.95  &  18.82 $\pm$ 1.85 &  \textbf{29.59} $\pm$ 2.22     & \multirow{2}{*}{TV, U-Net, E2E-Varnet}   \\
                             & &                      & SSIM          &  0.510 $\pm$ 0.093  &  0.537 $\pm$ 0.077  &  0.548 $\pm$ 0.050     &         \textbf{0.702} $\pm$ 0.028 & \\
\hline

\end{tabular}
\end{adjustbox}
\caption{{Quantitative metrics of multi-coil parallel imaging}. Numbers are presented as mean value $\pm$ unit standard deviation. Numbers in bold face indicate the best metric out of all the methods. $p$-values indicate results from RM-ANOVA. Proposed method statistically relevant against methods listed in the last column in terms of PSNR. $^*$: Bonferroni corrected.}
\label{tbl:quantitative_metric_multi}
\end{table*}

\subsection{Real-valued Simulation study}
\label{sec:result_real}

First, we show the results of the simulation study using retrospectively under-sampled fastMRI real-valued data of size 320$\times$320 in Fig.~\ref{fig:real_results}. In the first row, we see reconstructions from 2D $\times 8$ Gaussian random sampling, which is the sub-sampling pattern that induces the least aliasing artifact. Here, we see that the proposed performs nearly perfect reconstruction, where we see virtually no structural difference even in the zoomed-in image. Reconstruction through total variation (TV) induces cartoon-like artifacts, and cannot remove the overall foggy artifacts across the whole image. U-Net and DuDoRNet produce blurry reconstructions that are suboptimal, which indicates that they cannot adapt to different sampling patterns that were not seen from the training process. 

In the second row, we compare the reconstructions from acceleration using $\times 4$ uniform random sampling with 4\% of the phase encoding lines kept as the autocalibrating signal (ACS) region, producing moderate aliasing artifacts. Even at this level, our method is able to provide accurate reconstruction, capturing most of the high-frequency details of the ground truth. In this acceleration factor, TV is not able to completely remove the aliasing artifacts. The averaging effect of supervised reconstruction now becomes more clear, often omitting the important details of the scan. Reconstruction with DuDoRNet also suffers from leftover aliasing artifacts in this regime, even though this is just a slight difference in the sampling pattern.

The third row shows $\times 8$ Gaussian random sampling (4\% ACS region). TV only partially removes the artifacts, while introducing quite an amount of cartoon-like artifacts. Supervised U-Net and DuDoRNet smooths away a lot of details, producing unrealistic texture. Surprisingly, even at this level of acceleration factor, our method provides a fairly accurate reconstruction, matching most of the details from the ground truth.

In the last row, we see the most aggressive under-sampling factor - $\times 15$ variable density poisson disk under-sampling. Thanks to the evenly spread out $k$-space samples, this sampling scheme is considered one of the state-of-the-art amongst the various sampling patterns, which produce the least artifacts given the same budjet~\citep{dwork2021fast}. Here, similar to what was seen with 2D Gaussian sampling, only the proposed method is able to produce high-fidelity image reconstruction. Other methods still suffer from severe artifacts, with blurred-out details.

We provide a thorough comparison of quantitative metrics on the test set in Table~\ref{tbl:quantitative_metric_single}. On all the different sampling patterns, our method significantly outperforms the comparison methods and is on par with the state-of-the-art DuDoRNet~\citep{zhou2020dudornet} in Gaussian 1D sampling pattern. Results of RM-ANOVA also indicates that the superiority of the proposed method is statistically significant against all other methods in most cases.

It is worth mentioning the overall difference between the reconstructions through supervised methods (i.e., U-Net, DuDoRNet), and the proposed method. We observe that the reconstructions using supervised U-Net become blurrier as we push the acceleration factor to higher values. This is a widely known effect for models that are trained with supervision using L1 loss, L2 loss, etc~\citep{ledig2017photo}. Typically, this behavior stems from the models collapsing to a single mode of the distribution, whereas the actual distribution is highly multi-modal. On the contrary, our method does not show such an effect, visiting the modes of high probability in each sample. For that matter, our method is able to 1) reconstruct high-frequency details even at high acceleration factors, and 2) quantify uncertainty, as we discuss in-depth in Section~\ref{sec:uncertainty}

\subsection{Complex-valued Single-coil Reconstruction}
\label{sec:result_single}

In Fig.~\ref{fig:singlecoil_results}, we illustrate the reconstruction results of the proposed method using Algorithm~\ref{alg:PC-POCS_complex}, along with the baseline comparisons. In the first row, we have $\times 8$ 2D Gaussian random sampling acceleration. The proposed method performs a very accurate reconstruction, without inducing any additional blurriness from the reconstruction process. TV, supervised U-Net, and DuDoRNet fall largely behind the proposed method both in terms of perceptual quality and quantitative metrics.

In the second row, we have $\times 4$ 1D Gaussian random sampling (8\% ACS region) acceleration. Again, the proposed method reconstructs the aliased image with high accuracy and has higher quality than the comparison methods.

The third row shows $\times 8$ 1D uniform random sampling (4\% ACS region). The proposed method is still able to reconstruct the aliased image with virtually no degradation in the high-frequency details. On the other hand, supervised U-Net reconstruction produces horizontal strip artifacts, hampering the visual quality. DuDoRNet does not remove the aliasing artifacts completely. TV hardly produces satisfactory results. The final row shows reconstructions with $\times$15 VD poisson sampling. Due to the bias toward 1D sampling patterns, both U-Net and DuDoRNet produce unsatisfactory results with washed-out details. Our method is able to preserve sharp edges and texture, clearly depicting the anatomical structure. 
Quantitative metrics in Table~\ref{tbl:quantitative_metric_single} also confirm the superiority.

As mentioned earlier, all the results depicted in Fig.~\ref{fig:singlecoil_results} were generated using the single score function, trained with DICOM images only. This means that the model has never seen complex-valued data before, and yet the quality of reconstruction is surprisingly high. This is an important advantage of our model over conventional DL methods. Most DL methods require raw $k$-space data to train the network, and this is hard to achieve since most raw data are discarded after the scan~\citep{zbontar2018fastmri}. In contrast, DICOM images are relatively much easier to collect, and this enables practitioners to collect a large database for training.

\subsection{Complex-valued Multi-coil Reconstruction}
\label{sec:result_multi}

\begin{figure*}[!hbt]
    \centering\includegraphics[width=18cm]{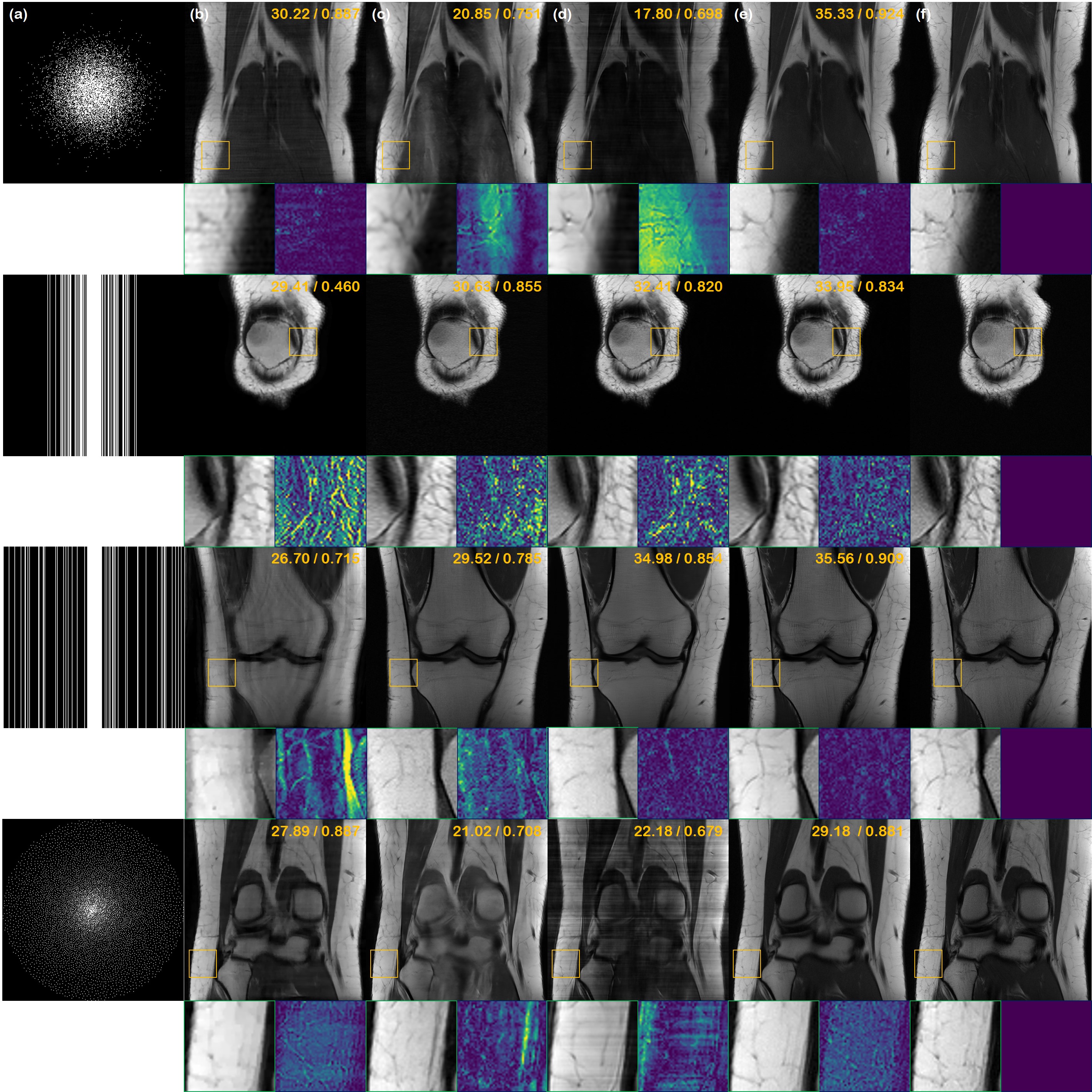}
    \caption{
    Multi-coil reconstruction results. (a) Sub-sampling mask used to generate under-sampled image, (b) TV, (c) supervised learning (U-Net), (d) E2E-varnet~\citep{sriram2020end}, (e) the proposed method,  and (f) the ground truth. 1$^{\text{st}}$ row: 2D $\times$8 Gaussian random sampling, 2$^{\text{nd}}$ row: 1D $\times$4 Gaussian random sampling, 3$^{\text{rd}}$ row: 1D $\times$ 4 uniform random sampling, 4$^{\text{th}}$ row: $\times$8 variable density poisson disk sampling. Green box: Zoom in version of the indicated yellow box, Blue box: Difference magnitude of the inset (in \code{Viridis} colormap). Yellow numbers in the upper right corner indiate PSNR [db], and SSIM, respectively.}
	\label{fig:multicoil_results}
\end{figure*}

We compare the results of PI reconstruction in Fig.~\ref{fig:multicoil_results}. Consistent with the results from the prior sections, our method significantly outperforms other methods and is the only method that produces high fidelity reconstructions regardless of the sampling pattern. This can also be seen in Table~\ref{tbl:quantitative_metric_multi}, where we see that our method is on par with the SOTA E2E-varnet on 1D sampling patterns, while significantly outperforming all the methods in 2D sampling patterns.

The bias towards 1D sampling pattern for the deep-learning based method is much larger in the multi-coil case. This can be seen in the severe artifacts that can be seen in reconstructions via E2E-varnet (Fig.~\ref{fig:multicoil_results} (d)). We conjecture that this is because the sensitivity map estimation module of E2E-varnet fails dramatically with 2D sampling patterns, leading to horizontal streaking artifacts. Furthermore, although U-Net does not explicitly deal with sensitivity maps, the bias towards certain sampling patterns seems to be greater than in the single-coil experiments, showing reconstructions with heavy artifacts. On the other hand, our score-based method is agnostic to sampling patterns, clearly outperforming the comparison methods in all cases.

However, in the case of PI, we do observe some potential limitations of our method. Namely, since we consider stochastic samples from the conditional distribution for each coil, and then take the SSOS of the coil images at the last step to form the reconstruction, we recognize a small amount of averaging effect compared to single-coil reconstruction cases. Nevertheless, we emphasize that our method was not trained with multi-coil data, and is yet able to reconstruct PI data with state-of-the-art performance. Our method works with any arbitrary number of coils, without the need to calibrate sensitivity maps, thereby enabling a practical application to any type of scan.

\subsection{Pathology detection}
\label{sec:pathology}

\begin{figure*}[!hbt]
    \centering\includegraphics[width=18cm]{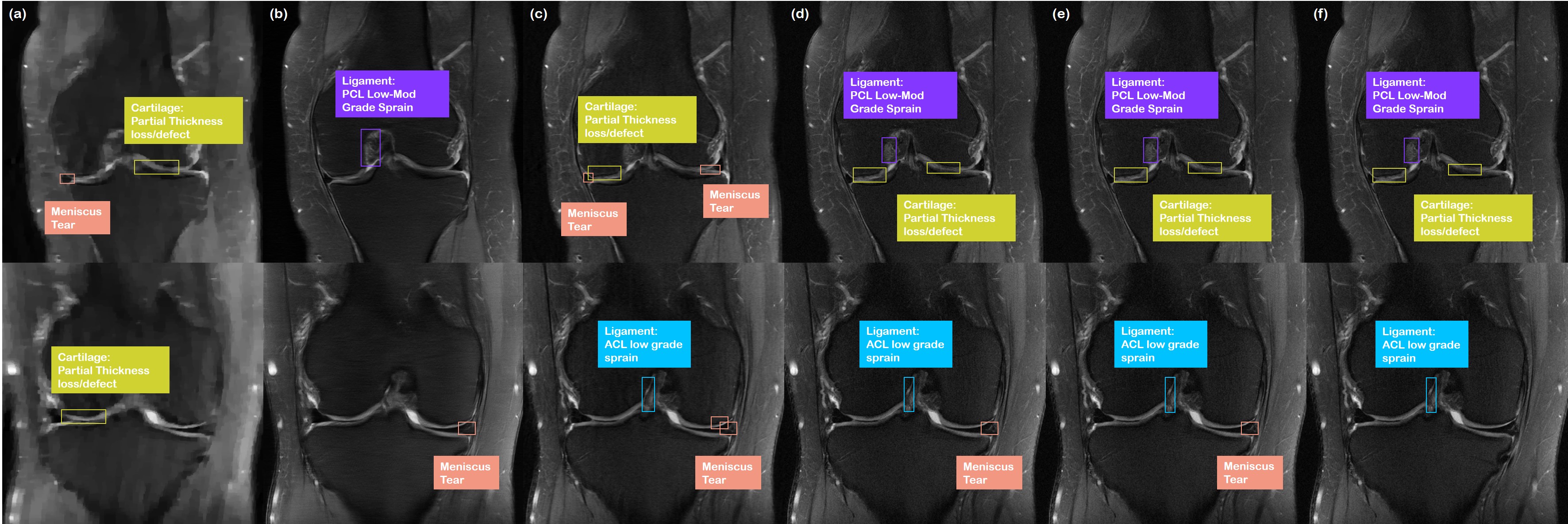}
    \caption{Results of pathology detection. Detection using (a) TV reconstruction, (b) supervised U-Net, (c) DuDoRNet, (d) proposed method, (e) fully-sampled images. Ground-truth label for the pathologies are shown in (f). (Yellow green box): Cartilage partial thickness loss/defect, (Pink box): Meniscus tear, (Purple box): Ligament PCL Low-Mod grade sprain, (Skyblue box): Ligament ACL low grade sprain.}
	\label{fig:yolo_results}
\end{figure*}

\begin{figure}[!hbt]
    \centering\includegraphics[width=9cm]{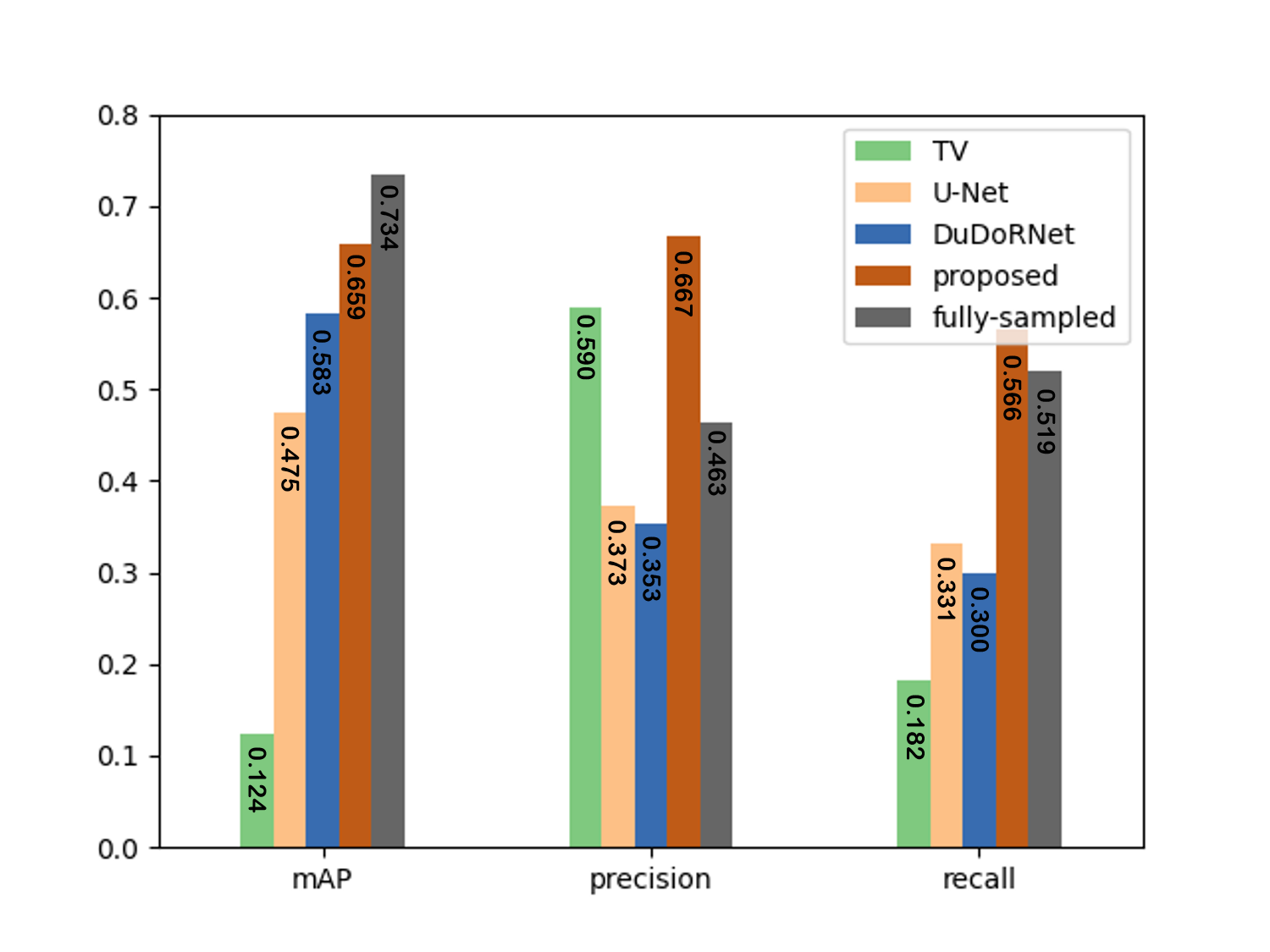}
    \vspace*{-0.9cm}
    \caption{Quantitative metrics of pathology detection.}
	\label{fig:yolo_results_bargraph}
\end{figure}

Overall results on the pathology detection task is illustrated in Fig.~\ref{fig:yolo_results}, and the quantitative metric is shown in Fig.~\ref{fig:yolo_results_bargraph}. The images in Fig.~\ref{fig:yolo_results} (a-d) were reconstructed from $\times 4$ 1D Gaussian random under-sampling (8\% ACS region), (e) is the detection result using fully-sampled data, and (f) is the ground-truth bounding boxes along with the corresponding labels.

The patient scan in the first row of Fig.~\ref{fig:yolo_results}(f) shows grade 1 sprain on the posterior cruciate ligament (PCL). Moreover, both on the medial and the lateral side of the cartilage, partial-thickness loss/defect is observed. This is accurately captured in Fig.~\ref{fig:yolo_results} (e), where all three bounding boxes are correctly predicted. Reconstruction with the proposed method come as close, with virtually no difference with the fully-sampled reference (see Fig.~\ref{fig:yolo_results}(c)). Reconstruction with supervised U-Net performs much worse on this downstream task, missing the cartilage partial thickness loss/defect (Fig.~\ref{fig:yolo_results}(b)). DuDoRNet reconstruction gets only one of the cartilage partial thickness loss/defect, while falsely detecting meniscus tears. As expected, diagnosis with TV reconstruction performs even worse, falsely predicting meniscus tear on both medial/lateral sides.

The patient scan in the second row of Fig.~\ref{fig:yolo_results}(f) show a grade 1 sprain on the anterior cruciate ligament (ACL). On both fully-sampled data, and reconstruction using the proposed method in Fig.~\ref{fig:yolo_results}(e),(d), respectively, the model correctly predicts ACL low-grade sprain, with small differences in the size of the bounding box. Both models also falsely predict meniscus tear, which seems reasonable given that the model is not perfect. Contrarily, reconstruction using supervised U-Net in Fig.~\ref{fig:yolo_results}(b) does not capture ACL low-grade sprain, and only falsely estimates meniscus tear. DuDoRNet comes similar to fully-sampled data, but is confused at the exact location of the meniscus tear. TV reconstruction in Fig.~\ref{fig:yolo_results}(a) only falsely predicts cartilage loss/defect, missing the true lesion.

As can be seen from standard metrics in Fig.~\ref{fig:yolo_results_bargraph}, the mAP metric using the fully-sampled data reaches 0.754, which serves as the upper bound for the detection model. Detection using the proposed reconstructions performs on par, or sometimes even better than the fully-sampled reference. Since the model is imperfect and the test dataset size is relatively small, this does not mean that the proposed method is better than the fully-sampled data. However, it does mean that reconstruction via the proposed method does not hamper the diagnostic capability, which is important in clinical practice. On the other hand, the comparison methods do limit the diagnostic capability, and thus should be questioned before usage in clinical practice. For further statistical analysis, see Appendix.

\subsection{Quantifying uncertainty of the prediction}
\label{sec:uncertainty}

\begin{figure}[!hbt]
    \centering\includegraphics[width=8cm]{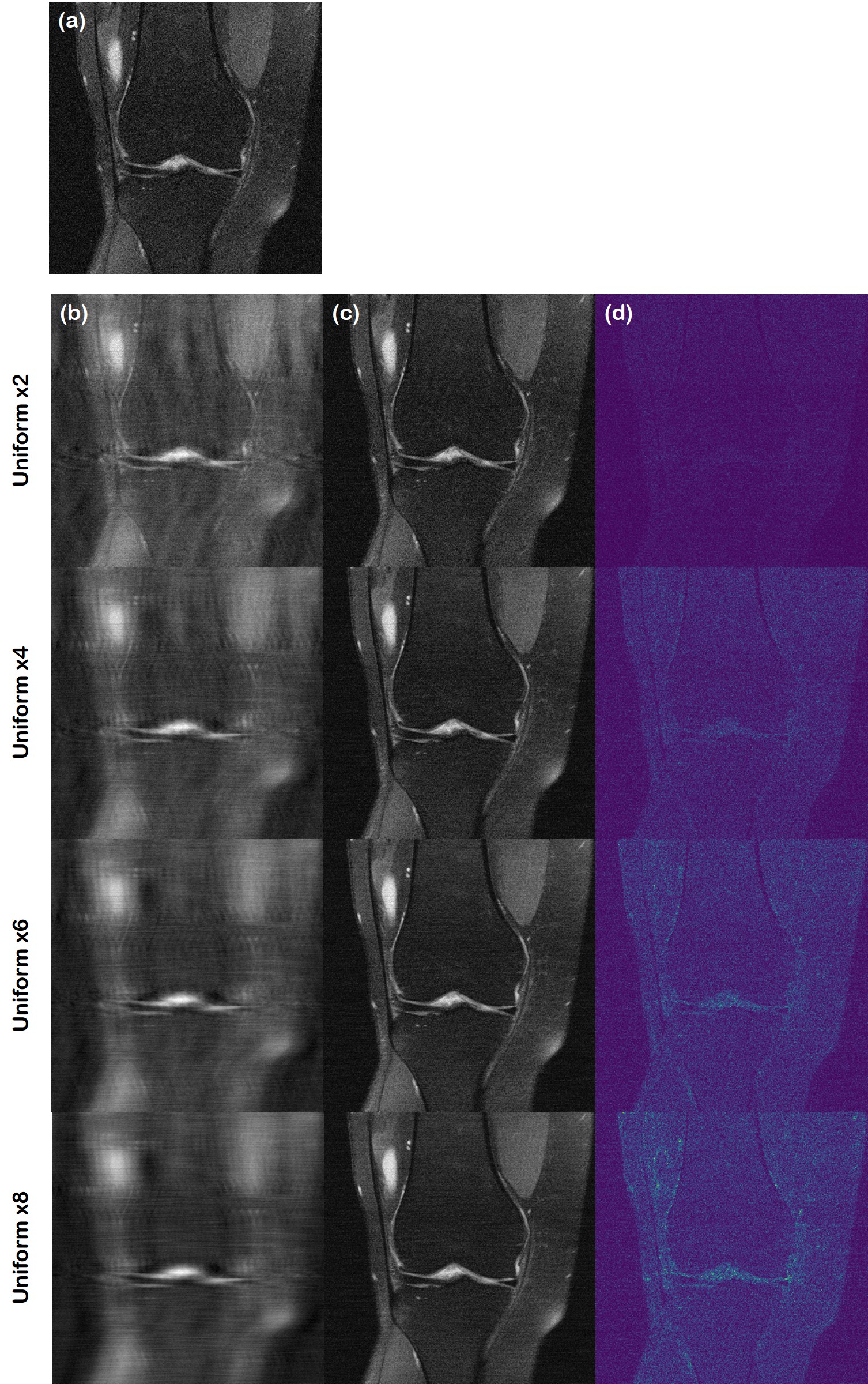}
    \caption{Quantifying the uncertainty of reconstruction. (a) Ground truth, (b) aliased image from sub-sampling, (c) mean of the reconstruction, (d) standard deviation of the samples: range is set to [0, 0.02] (on \code{Viridis} colormap). From the 1$^\text{st}$ row to the 4$^\text{th}$ row, the acceleration factor grows from $\times 2$ to $\times 8$.}
	\label{fig:uncertainty}
\end{figure}

Our method is a generative algorithm, with two sources of stochasticity. First, the sample starts from a randomly sampled vector $\xb_N$. Second, both predictor and corrector steps involve sampling random noise vectors and adding them to the estimate. Therefore, the iterative procedure of the proposed algorithms typically converges to different outcomes. 

Due to this generative nature, we can run multiple reconstructions in parallel, and quantify the uncertainty of the prediction, as depicted in Fig.~\ref{fig:uncertainty}. Here, the ground truth, and the aliased images are shown in Fig.~\ref{fig:uncertainty} (a),(b), respectively. For the experiment, we take a batch size of 8, and run the reconstruction in parallel. The mean value of the reconstruction is shown in Fig.~\ref{fig:uncertainty} (c), and the pixel-wise standard deviation values are shown in Fig.~\ref{fig:uncertainty} (d). At low acceleration factors ($\times 2$), we see very little variation between the different reconstructions. This indicates high confidence of the model, and hence we can conclude that the reconstruction is exact in all parts of the image. As the acceleration factor is increased, and the degree of aliasing artifacts become more severe, we see that the uncertainty increase in specific regions. Potentially, this measure of uncertainty can inform the practitioners on how much they should rely on the reconstruction, thereby deciding whether to use a different diagnostic tool. We additionally provide specific realizations, rather than the mean of the reconstruction, at each acceleration factor in supplementary video 1.

\subsection{Reconstruction out-of-distribution (OOD) data}

\begin{figure*}[!hbt]
    \centering\includegraphics[width=16cm]{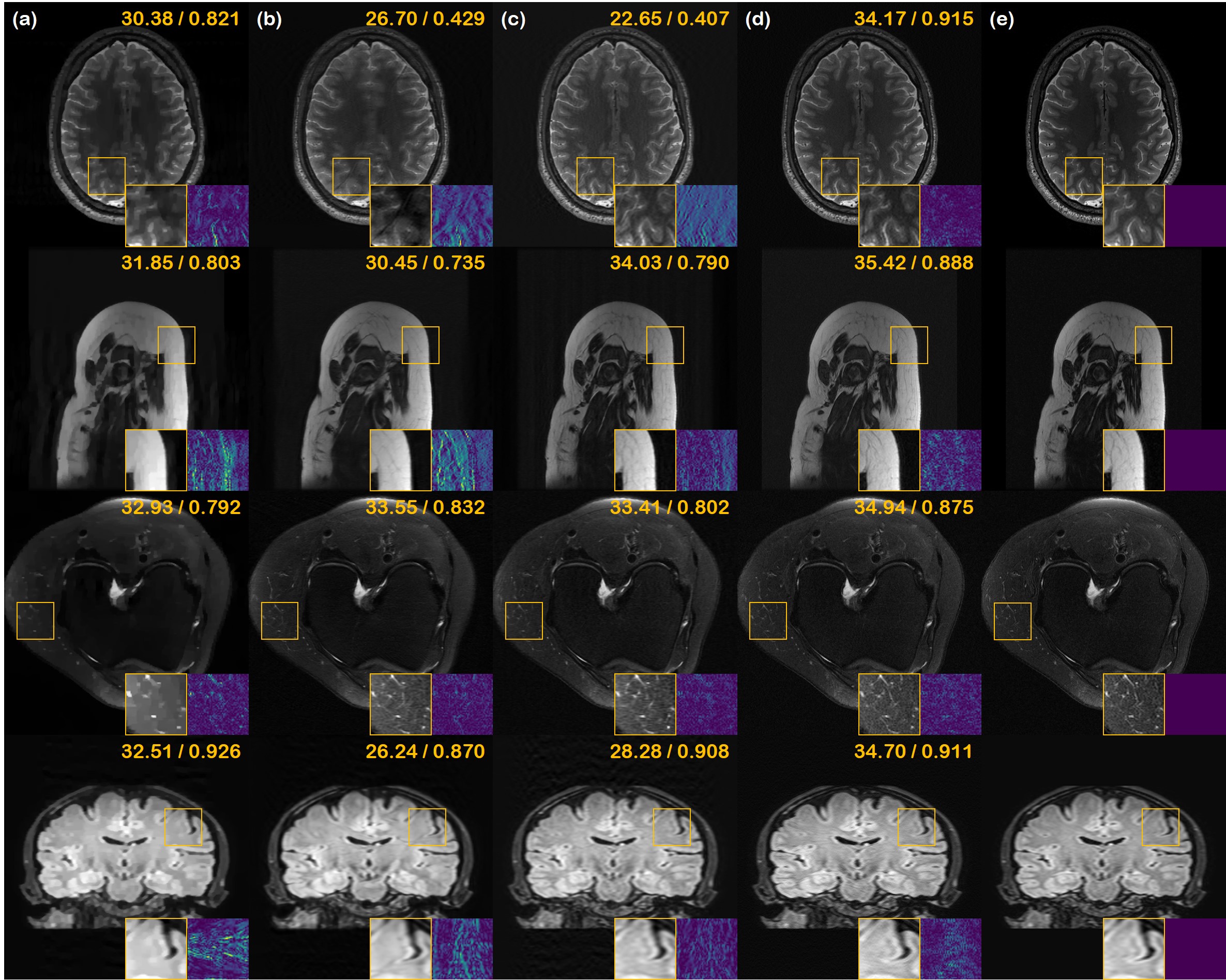}
    \caption{Reconstruction results of various anatomic structure/contrast. (a) TV, (b) U-Net, (c) DuDoRNet, (d) proposed method, (e) ground truth. 1$^{st}$ row: HCP axial brain scan, 2$^{nd}$ row: MRI scan of left leg collected from~\citep{mridata}, 3$^{rd}$ row: axial knee scan from~\citep{mridata}, 4$^{th}$ row: MASSIVE coronal brain scan. 1D Gaussian $\times4$ under-sampling was performed.}
	\label{fig:va}
\end{figure*}

In Fig.~\ref{fig:va}, we present reconstruction results, which are heavily out of distribution. Note that our score function has only learnt the distribution of proton density (PD) / proton density fat suppression (PDFS) coronal knee scans. Nevertheless, we observe that with the proposed method, we are able to achieve high fidelity reconstructions regardless of the anatomy and contrast. While other methods such as U-Net and DuDoRNet generalizes to a certain extent, we can clearly observe leftover aliasing artifacts. On the other hand, the proposed method clearly outperforms all the other methods with minimal residuals and sharp contrast. We recently found that
 this observation was also made independently in~\citep{jalal2021robust}, where the authors partially proved that posterior sampling is indeed highly robust to distribution shifts. This property is indeed very advantageous in real-world settings, since one may be able to use a single neural network regardless of the specific anatomy and contrast. For further experimental results of different anatomy, please see Fig.~\ref{fig:va_supp1} and Fig.~\ref{fig:va_supp2} in the Appendix.

\section{Discussions}
\label{sec:discussion}

\subsection{Speeding up inference}
\label{sec:speed}

\begin{figure}[!hbt]
    \centering\includegraphics[width=8cm]{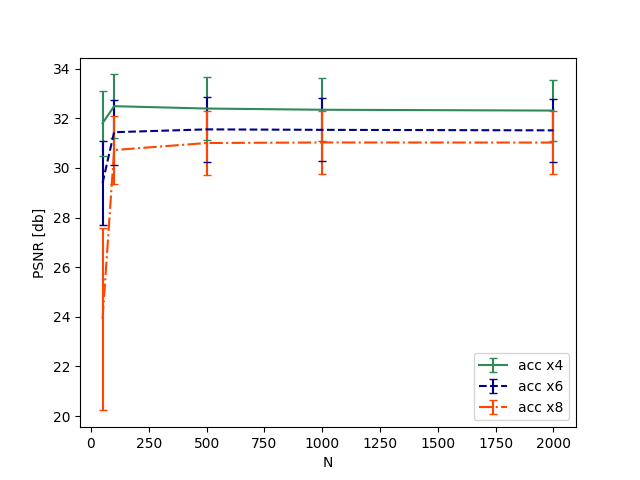}
   \vspace*{-0.5cm}
    \caption{Tradeoff between number of iterations vs. PSNR [db]. Errorbars indicate unit standard deviation. Sampling scheme used for this ablation study is 1D Gaussian random under-sampling.}
	\label{fig:acceleration}
\end{figure}

One obvious limitation of using score-based diffusion models for image reconstruction is the time required for inference. As stated in Section~\ref{sec:imple_details}, it requires about 10 minutes of inference time for using $N = 2000$ discretization steps. A naïve way for faster inference is to reduce the number of discretization steps, and we provide the trade-off between image quality vs. steps in Fig.~\ref{fig:acceleration}. Here, we observe that naïvely interleaving the discretization steps work quite well, with minimal compromise in image quality. This is especially the case of low acceleration factors (e.g. $\times 4$), where we achieve highly accurate reconstruction only with 50 iterations. As the acceleration factor gets aggressive, our method typically requires more iterations for maximal performance. However, from the figure, we see that the performance caps at about $N = 500$. One could possibly adjust this as a hyperparameter on-the-fly, according to the degree of acceleration at hand.

One can also employ the state-of-the-art  acceleration strategy of diffusion models for inverse problems, called come-closer-diffuse-faster (CCDF), which was recently proposed in our companion work~\citep{chung2021come}. 
 Specifically, CCDF states that there is no need to use the full reverse diffusion procedure. Rather, one can 
start to apply reverse diffusion from a forward-diffused image from better an initialization
 to achieve reconstruction performance that is one par or {\em better}. 
 The resulting short partial time horizon significantly accelerates the reconstruction time.
 The strategy is backed by rigorous proofs using the theory of stochastic contraction. Interested readers are referred to~\cite{chung2021come}.

\begin{algorithm}[!t]
\caption{CCDF sampling \citep{chung2021come}}
\begin{algorithmic}[1]
\Require $\sb_{\theta}, N', M, \{\epsilon_i\}_{i=1}^{N'}$, $\{\sigma_i\}_{i=1}^{N'}, \xb_0$ \\
\textbf{Define} $A := \Pc_{\Omega} \Fc$
\State $\xb_{N'} \gets \xb_0 + \sigma_{N'}\zb$
\For{$i = N'-1:0$} \do \\
\State $\xb_i \gets \text{Predictor}(\xb_{i+1}, \sigma_i, \sigma_{i+1})$
\State $\xb_i \gets \operatorname{Re}(\xb_i + A^*(y - A\xb_i))$
\For{$j = 1:M$} \do \\
\State $\xb_i \gets \text{Corrector}(\xb_{i}, \sigma_i, \epsilon_i)$
\State $\xb_i \gets \operatorname{Re}(\xb_i + A^*(y - A\xb_i))$
\EndFor
\EndFor
\State \textbf{return} $\xb_0$
\end{algorithmic}\label{alg:PC-POCS_CCDF}
\end{algorithm}

More specifically,
one specifies a significantly short timestep $t_0 < T$ (in the discretized setting, this corresponds to $N' := t_0 N$). 
Then, the initial reconstruction $\xb_0$ is forward-diffused with the prescribed forward SDE in a single step. In the case of VE-SDE, this corresponds to
\begin{equation*}
    \xb_{N'} = \xb_0 + \sigma_{N'}\zb,\quad \zb \sim \Nc(\mathbf{0}, \Ib).
\end{equation*}
One can then follow the reverse SDE for $t \in [0,t_0]$ running backwards in time as in Algorithms~\ref{alg:PC-POCS},\ref{alg:PC-POCS_complex},\ref{alg:PC-POCS_GRAPPA},\ref{alg:PC-POCS_hybrid}.

For simplicity, here we present CCDF adopted to Algorithm~\ref{alg:PC-POCS}, which is shown in Algorithm~\ref{alg:PC-POCS_CCDF}. For the initialization of $\xb_0$, we resort to U-Net. With this simple combination of the pre-trained feed forward neural network, we observe that it is possible to use as little as {\em 40 iterations} (corresponding to 0.02$\times$ NFEs\footnote{short for Neural Function Evaluations}) with similar or better performance, as reported in Table~\ref{tab:CCDF}.

\begin{table}[!hbt]
    \centering
    \resizebox{0.45\textwidth}{!}{
    \begin{tabular}{c|ccccc}
    \hline
    method & TV & U-Net & DuDoRNet & \thead{proposed \\ (2000)} & \thead{{CCDF} \\ ({40})} \\ \hline\hline
    $\times$ 4 & 30.77 & 32.85 & 33.01 & 33.32 & \textbf{34.11} \\
    $\times$ 8 & 28.87 & 30.81 & 30.46 & 30.94 & \textbf{32.08} \\ \hline
    \end{tabular}
    }
    \caption{Effect of acceleration using the CCDF~\cite{chung2021come} strategy to the proposed method.
    PSNR($\uparrow$) on Gaussian 1D under-sampling masks with different acceleration factors are reported. Number in parenthesis indicate the number of iterations used. Numbers in boldface indicate the best results among the rows.}
    \label{tab:CCDF}
\end{table}

\subsection{Conditional generation with diffusion models}

Score-based diffusion models are now one of the most popular methods in the context of image synthesis, matching the image fidelity of state-of-the-art GANs~\citep{dhariwal2021diffusion}, and achieving the state-of-the-art log-likelihood on various datasets~\citep{kingma2021variational, kim2021score}. The interest in using these models for conditional image generation is also rising. \cite{song2019generative} first proposed to use score models trained with discrete de-noising score matching for image inpainting. This was further developed in \citep{song2020score} for image colorization, and class-conditional image synthesis, using continuous-time score models. The same group published a work for image editing~\citep{meng2021sdedit} using VE-SDEs, which uses a similar algorithm to image inpainting used in \citep{song2019generative, song2020score}. ILVR~\citep{choi2021ilvr} adopts diffusion models~\citep{ho2020denoising} for image super-resolution and image translation. All these works require training of a score model irrelevant to the actual objective task, and are thus flexible. Nevertheless, all the prior works have focused on applications where the condition also stays in the image domain, which makes the problem easier to solve. Our method adds further flexibility by showing that conditions can be applied in measurement domains that are not necessarily in the same image domain.

We are aware of one prior work which used denoising score matching for MRI reconstruction~\citep{ramzi2020denoising}. The authors of \citep{ramzi2020denoising} use amortized residual de-noising autoencoder (AR-DAE) score matching loss~\citep{lim2020ar} to train the score function, then uses annealed Hamiltonian MC~\citep{neal2011mcmc} to perform reconstruction from measurement. However, \citep{ramzi2020denoising} reported that their method falls behind supervised learning approaches by a large margin, especially when considering single samples. Moreover, the training methodology in~\citep{ramzi2020denoising} targets separate channel complex-valued data, which limits their application. Our method, on the other hand, beats neural networks trained with supervision, and requires only the magnitude images for training. It is also notable that the proposed method is applicable to PI.

After the submission of this paper, 
we found two independent  works that are closely related to the proposed work.
\cite{jalal2021robust} proposes to use score-based generative models to train a score function that is similar to ours, and sample by taking annealed Langevin dynamics (ALD)~\citep{song2019generative}  together with the gradient information with respect to the data fidelity term by assuming Gaussian measurement noise. As was also shown in our work, \cite{jalal2021robust} illustrates the robustness of using a score-based generative model for reconstruction over different sub-sampling patterns and diverse anatomy.
The largest difference between us and \cite{jalal2021robust} comes from the fact that our method only requires DICOM images for training the score function. This is in stark contrast with \cite{jalal2021robust}, as they require fully-sampled $k$-space data for training the score function. Furthermore, our work is based on the continuous version of score matching~\citep{song2020score}, whereas the work of \cite{jalal2021robust} is based on a discrete version~\citep{song2020improved}. It is also worth mentioning that we use an advanced sampler (PC), and a more efficient network architecture, which was shown to improve the performance of generative modeling by a large margin~\citep{song2020score}. Finally, \cite{jalal2021robust} introduces {\em annealing} the data fidelity gradient terms which require specifying the variance schedule per each noise scale. Our method does not have additional hyper-parameters, and is hence easier to implement.

\cite{song2022solving} is perhaps the most related to our work, in that the authors also propose to use VE-SDE of \citep{song2020score}, and they use the same network architecture together with the PC sampler, as in our work.
The crucial difference of our work from \cite{song2022solving} is that we derive solvers that are capable of reconstructing complex-valued data, and also multi-coil data. \cite{song2022solving} only focused on solving simulated reconstructions from real-valued images, which limits the practicality. Moreover, the data fidelity imposing step differs slightly from our work.

\subsection{Energy based models}

Energy-based models (EBMs) are non-normalized probabilistic models, which have the advantage of circumventing the need to compute the normalizing constant (i.e. the partition function)~\citep{song2021train}. Our work relies on denoising score matching to estimate the score, which also belongs to the category of EBMs. There are, of course, other ways to train EBMs, and one of the most widely known methods other than score matching, is contrastive divergence (CD)~\citep{carreira2005contrastive}.

Concurrent to our work, an MR acceleration algorithm utilizing CD was proposed~\citep{guan2021mri}. This work grounds their method on an EBM, trained with persistent contrastive divergence (PCD), which is a variant of CD using sample buffers~\citep{yilun2019implicit}. Specifically, a parameterized energy function $E_\theta(\cdot)$ is trained such that it takes low values when the input to the function is likely to be in the data distribution, and high values when this is not the case. Once the energy function is trained, one can generate unconditional samples via MCMC, or apply data consistency projection in between the MCMC update steps to sample from a conditional distribution.

This procedure is in fact similar to our method, with the trained functions forming the relationship of $\nabla_\xb \log p_\theta (\xb) = -\nabla_\xb E_\theta (\xb)$. However, we note two key differences of the proposed method from \citep{guan2021mri}. First, in order to train $E_\theta$ with CD, one has to produce negative samples with MCMC at every iteration of training, which is non-trivial and costly. Several heuristics need to be applied to make the algorithm work. In contrast, the training of our score function is much more straightforward and robust, breaking down to an explicit form of the loss function, as formulated in Eq.~\eqref{eq:score_cost_VESDE}. Second, our method only requires magnitude (DICOM) images for training, whereas \citep{guan2021mri} requires raw data. 

\subsection{Generative models for inverse problems}
\label{sec:gen_ip}

Before the recent surge of diffusion models, there have been several approaches to use generative models, especially generative adversarial networks (GAN)~\citep{goodfellow2014generative} as priors for solving inverse problems~\citep{marinescu2020bayesian, asim2020invertible}. These methods offer improved flexibility over supervised models which are trained for specific problems. Rather, one can leverage a well-trained generator $G_\varphi$, which was trained without the knowledge of the forward physics. In essence, in order to solve inverse problems with GAN prior, one would typically optimize for the following:
\begin{equation}
\label{eq:GANprior1}
    \min_\zb \|AG_\varphi(\zb) - \yb\|,
\end{equation}
where $A$ and $\yb$ are as defined in eq.~\eqref{eq:forward_model}, and $\zb$ denotes the latent (noise) vector. This corresponds to finding the correct latent vector $\zb$ which minimizes the data fidelity. One can also try
\begin{equation}
\label{eq:GANprior2}
    \min_\varphi \|AG_\varphi(\zb) - \yb\|,
\end{equation}
which corresponds to tweaking the model parameters $\varphi$ so that the generator {\em adapts} to the forward physics of the problem. Once the optimization via \eqref{eq:GANprior1} or  \eqref{eq:GANprior2} is complete, a single forward pass through the generator $G_\varphi$ is sufficient for reconstruction. Unfortunately, there are several problems with these methods.

First, both the problems \eqref{eq:GANprior1} and \eqref{eq:GANprior2} are nontrivial to solve, and require several heuristics such as using a sophisticated loss function~\citep{marinescu2020bayesian, asim2020invertible}. Considering that GANs themselves are also notoriously hard to train, methods that rely on GAN priors are relatively hard to reproduce.
Second, the final reconstruction step involves a {\em single} forward pass through $G_\varphi$. For highly ill-posed problems, it could be particularly hard to generate high-quality samples from this single pass, which could be a reason why it is hard to achieve a reconstruction with both high quality and data fidelity.

The proposed method, which proposes to use diffusion models instead of GANs, solves both problems. Diffusion models have a relatively well-defined loss and are thus easier to train. Moreover, the sampling procedure can easily be done with the most basic method of solving inverse problems. Further, with diffusion models, one can achieve fine-grained control since we iteratively refine our reconstruction.

\subsection{Broader Impact}

The proposed method can be readily applied to other problems in computational imaging, with well trained score function and the right modifications to the inference procedure. A single score function has already shown broad applicability: SR~\citep{choi2021ilvr, saharia2021image}, image reconstruction, and others. This could potentially shift the current paradigm of deep learning in biomedical imaging. For example, one can train a single score function for the imaging modality, and use it as a universal problem solver, given enough capacity. 

\subsection{Limitations}
\label{sec:limitations}

\begin{figure}[!hbt]
    \centering\includegraphics[width=8cm]{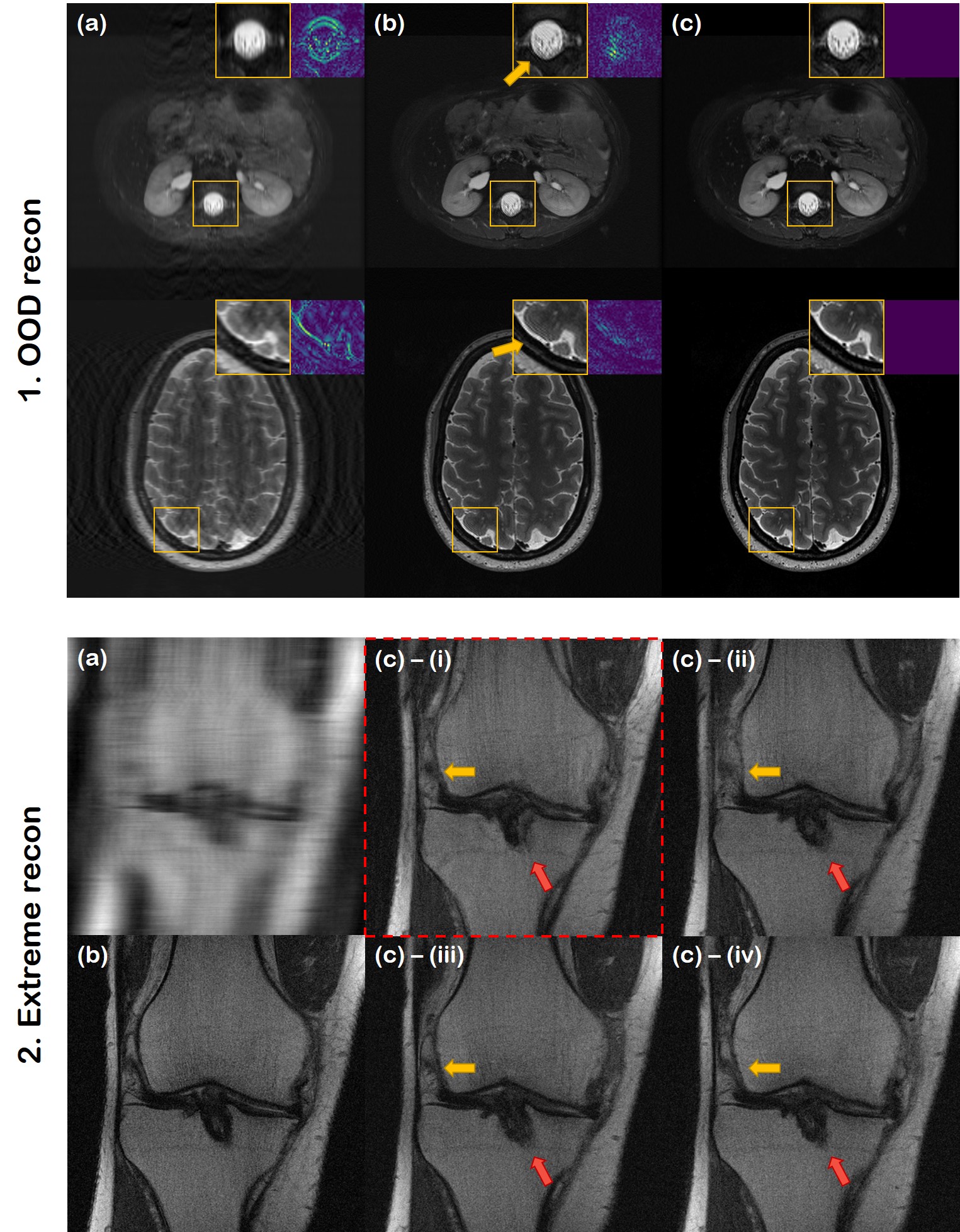}
    \caption{Limitations of the proposed method. 1. OOD reconstruction with 1D under-sampling pattern. (a) zero-filled, (b) proposed reconstruction, and (c) ground truth. Yellow arrows indicate the artifacts. 2. Extreme reconstruction (Uniform 1D $\times$15 acc.) (a) zero-filled, (b) ground truth, and (c) - (i~vi) posterior samples. Red dotted box indicates the posterior sample of the worst quality. Yellow and red arrows mark regions with high alteration.}
	\label{fig:limitation}
\end{figure}

For completeness, here we list two limitations of the current work. First, when we attempt to reconstruct OOD data with 1D under-sampling pattern, we sometimes observe mild aliasing-like artifacts in local edges. As illustrated in Fig.~\ref{fig:limitation} (1. OOD recon), the artifact is not significant. However, care should be taken when extending the proposed framework to OOD data, as the robustness will be compromised. We note that such an artifact is not observed in 2D sampling patterns, as was shown in Fig.~\ref{fig:va}.

Second, when pursuing extreme-condition reconstruction, as in Fig.~\ref{fig:limitation} (2. Extreme recon), we occasionally acquire results that are unsatisfactory (e.g. sample marked with the red dotted line). Moreover, we observe that the detailed structure has high variance within the posterior samples, due to the high ill-posedness. Hence, care should be taken when pushing the accelerating factor to very high values, by, for example, sampling multiple reconstructions and considering the uncertainty as was discussed in section~\ref{sec:uncertainty}.

\section{Conclusion}
\label{sec:conclusion}

In conclusion, we propose  a novel score-based reconstruction method for accelerated MRI. We train the gradient of the log data distribution with continuous-time denoising score matching using the magnitude data. Using the learned score as the prior, one can sample from the conditional distribution given the measurement by simply applying data consistency projection at every step. Our method produces reconstructions of high accuracy, whether it be single-coil, or multi-coil cases. Compared to prior arts, we show the superiority of our method both in terms of quality, and practicality.

We believe that our method opens up a new generation of methods for inverse problems in imaging. Direct application of our method to other venues to test the generality is an interesting direction of future research. Other than that, there still remain unanswered questions, for example, reducing the reconstruction speed gap between our method and feedforward neural network approaches. We expect that many interesting questions and answers will be actively discussed in the near future.

\section*{Acknowledgments}
This work was supported by the National Research Foundation (NRF) of Korea grant NRF-2020R1A2B5B03001980.

\section*{Appendix}

\begin{figure*}[!hbt]
    \centering\includegraphics[width=12cm]{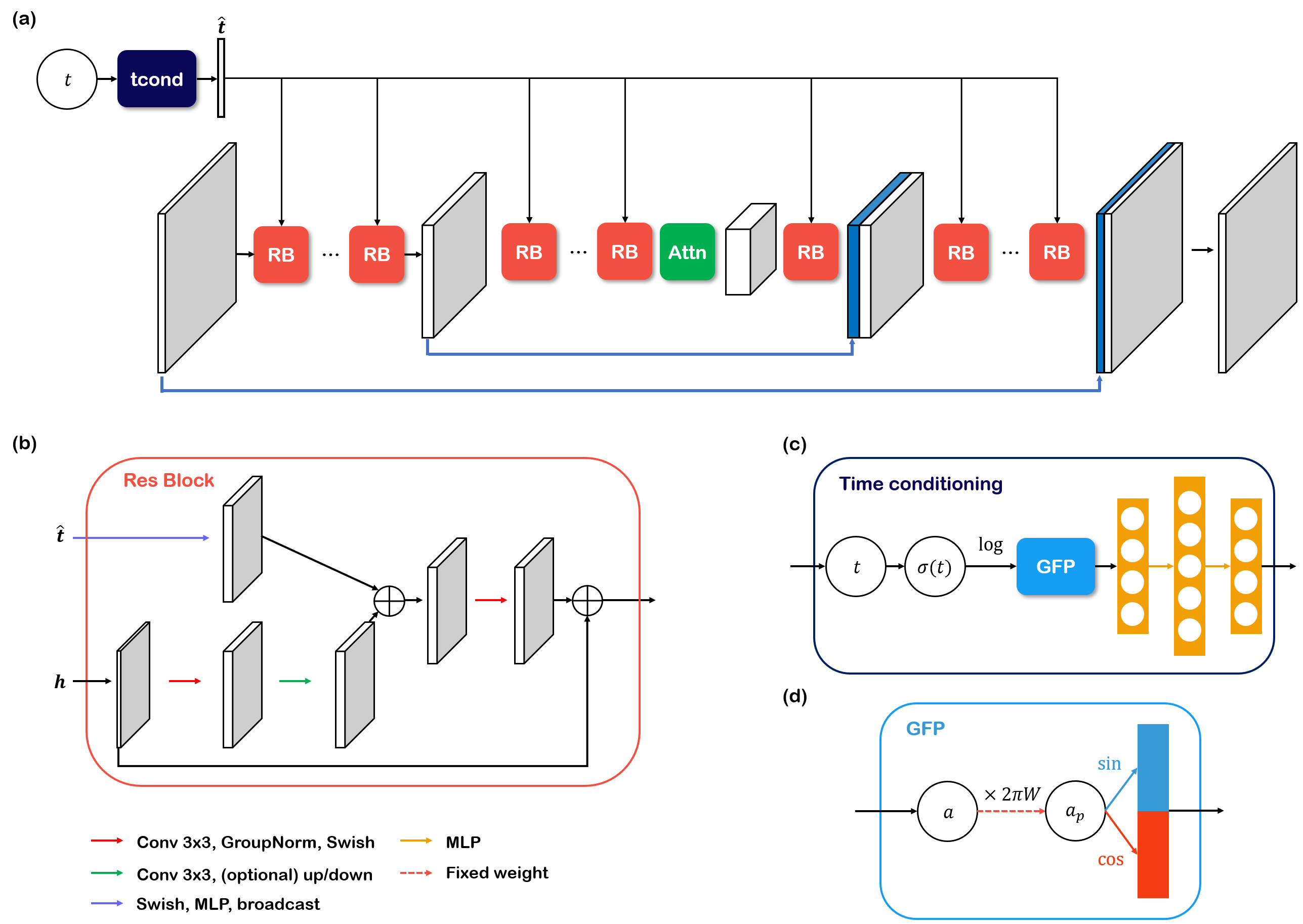}
    \caption{Illustration of the \code{ncsnpp} architecture used to model $s_\theta$. (a) Overall U-Net like structure, consisting of BigGAN residual blocks (RB), and attention blocks. Additional $t$ is incorporated with time conditioning block, with the output $\hat{\tb}$ affecting all the residual blocks. (b) Basic building block of \code{ncsnpp}. (c) Time conditioning module, which utilizes Gaussian Fourier Projection (GFP) module. (d) Detailed illustration of the GFP module.}
	\label{fig:net_arch}
\end{figure*}

\section*{Details of network architecture}

In this section, we elaborate on the \code{ncsnpp} network architecture used for constructing the score function $s_\theta$, especially focusing on the way the network incorporates time conditions. Overall illustration of the architecture is depicted in Fig.~\ref{fig:net_arch}. The network takes in as input, $\xb_t$, and $t$. Note that since we are solving for a continuous SDE, the additional time conditioning variable is important. For the time condition $t$, it passes through the time conditioning module, as shown in Fig.~\ref{fig:net_arch} (c), utilizing the Fourier features introduced in~\cite{tancik2020fourier}. In the figure, this is given as the Gaussian Fourier Projection (GFP) module, which can be mathematically written with:
\begin{align*}
    \text{GFP}(a) := &[\sin(2\pi w_1 a), \sin(2\pi w_2 a), \dots, \sin(2\pi w_d a), \\
    &\cos(2\pi w_1 a), \cos(2\pi w_2 a), \dots \cos(2\pi w_d a)],
\end{align*}
where the weights $w_1, w_2, \dots, w_d$ are non-trainable random initialized network parameters that are initialized with \code{torch.randn(d) * scale}, and we set \code{scale} to be 16. Subsequently, the embedding vector passes through a few layers of MLP to match the size of the channel dimension of input image features. Finally, the temporal feature vectors are broadcasted across the $H, W$ dimensions to be added to the image features via residual blocks at each level.

\section*{Statistical analysis of pathological detection}

\begin{figure*}[!hbt]
    \centering\includegraphics[width=15cm]{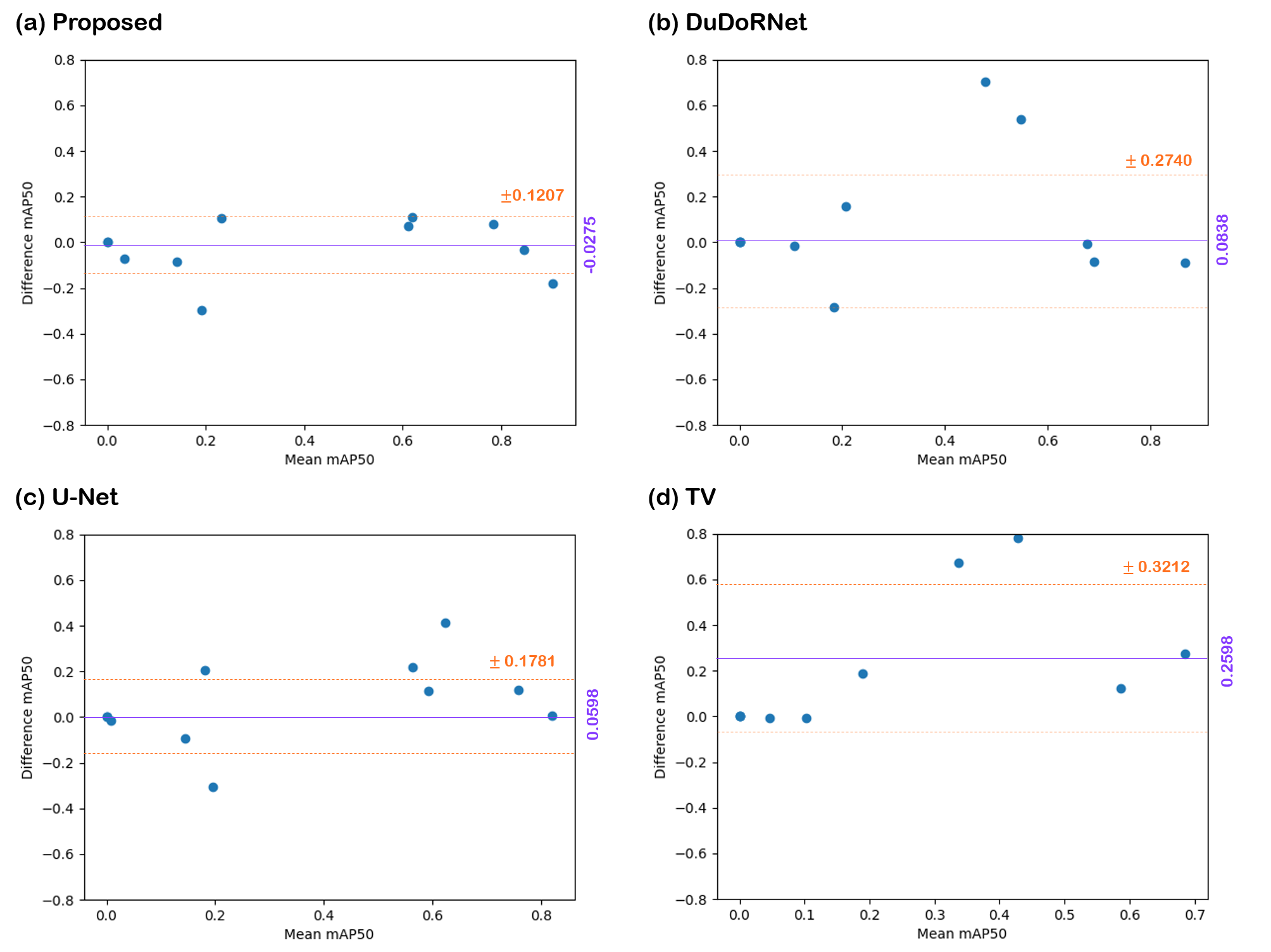}
    \caption{Bland-Altman plot of mAP score of pathologic detection with different methods. (a) Proposed method, (b) DuDoRNet~\citep{zhou2020dudornet}, (c) U-Net, (d) TV. $x$-axis represent the mean value, and the $y$-axis denote the difference from the result using fully-sampled data.}
	\label{fig:yolo_ba}
\end{figure*}

\begin{figure*}[!hbt]
    \centering\includegraphics[width=15cm]{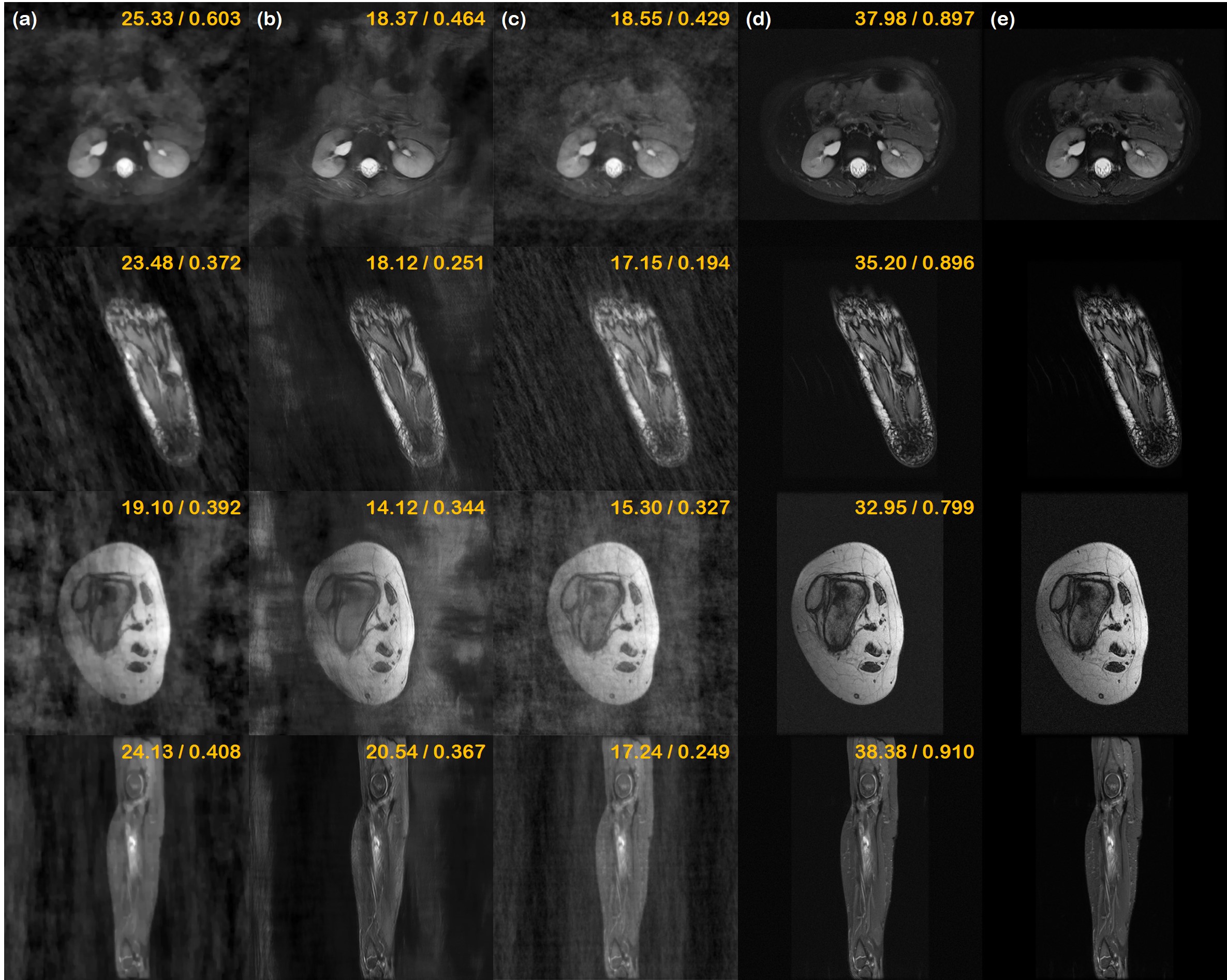}
    \caption{Reconstruction results of various anatomic structure/contrast. (a) TV, (b) U-Net, (c) DuDoRNet, (d) proposed method, (e) ground truth. All images are open-source data collected from sources indicated in Section~\ref{sec:va}. VD Poisson disk $\times8$ under-sampling was performed. All data were collected from~\cite{mridata}.
    1$^{st}$ row: cardiac MRI, 2$^{nd}$ row: foot axial scan, 3$^{rd}$ row: lower extremity axial scan, 4$^{th}$ row: LT bone.}
	\label{fig:va_supp1}
\end{figure*}

\begin{figure*}[!hbt]
    \centering\includegraphics[width=15cm]{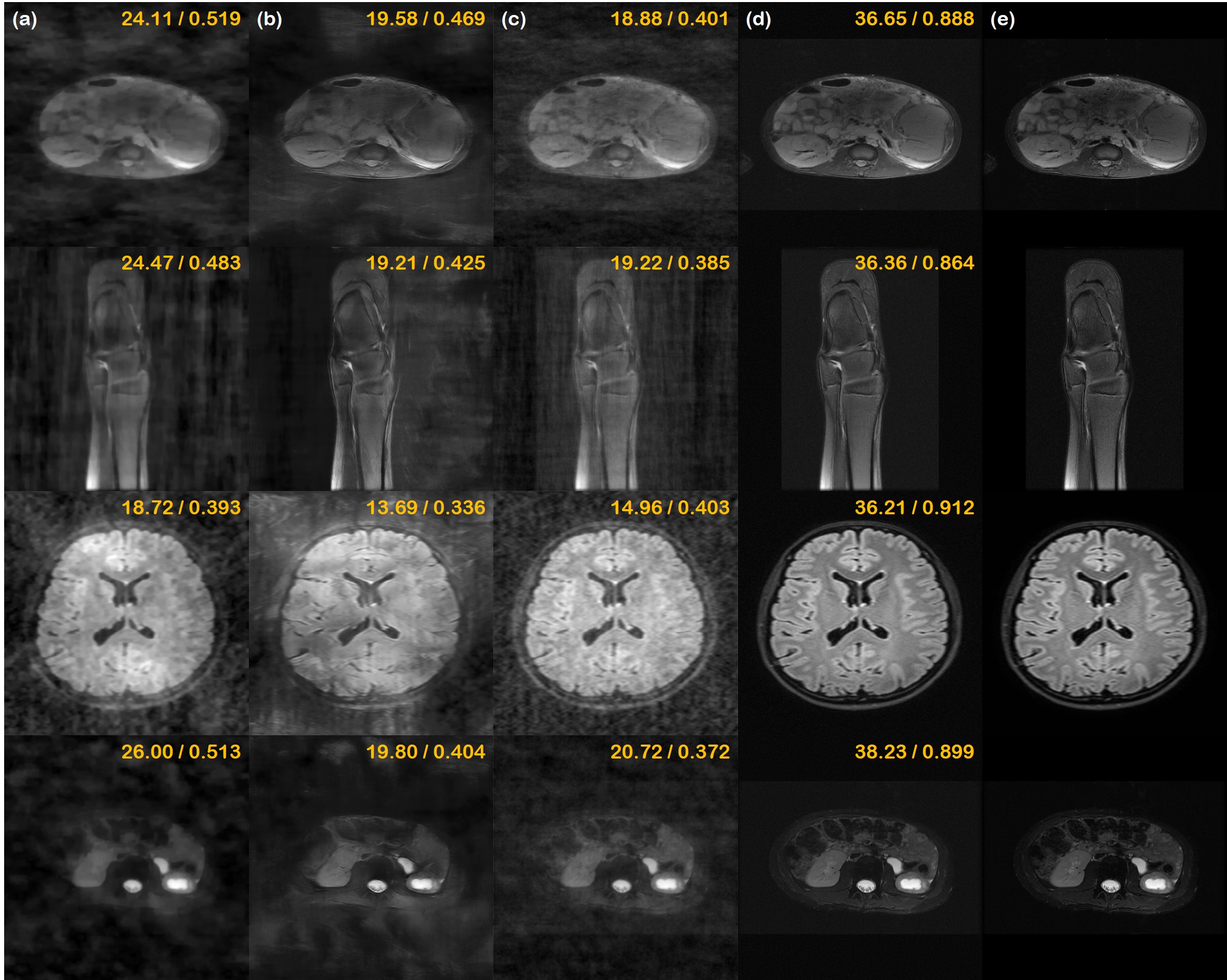}
    \caption{Reconstruction results of various anatomic structure/contrast. (a) TV, (b) U-Net, (c) DuDoRNet, (d) proposed method, (e) ground truth. All images are open-source data collected from sources indicated in Section~\ref{sec:va}. VD Poisson disk $\times8$ under-sampling was performed. The data in 1,2,4$^{th}$ row were collected from~\cite{mridata}, and the data in 3$^{rd}$ row was collected from MASSIVE.
    1$^{st}$ row: abdomen, 2$^{nd}$ row: ankle (foot), 3$^{rd}$ row: axial brain, 4$^{th}$ row: upper torso abdomen.}
	\label{fig:va_supp2}
\end{figure*}

Bland-Altman plot of the pathologic detection task is shown in Fig.~\ref{fig:yolo_ba}. For individual data points, we resort to the mAP50 score for each class.
With the plots, we can check for the agreement of each reconstruction method with the fully sampled data. Consistent with the observations that were made in the main text, we see that the most agreeing method is the proposed method, which shows the least variance in the difference. DuDoRNet and UNet both have higher variance, and TV is radically different from the fully sampled data.

\section*{Additional results}

Additional reconstruction results of different anatomies are presented in Fig.~\ref{fig:va_supp1} and Fig.~\ref{fig:va_supp2}.

\bibliographystyle{model2-names.bst}\biboptions{authoryear}
\bibliography{refs}

\end{document}